\documentclass[useAMS,usenatbib]{mn2e}
\usepackage{txfonts}
\usepackage{graphicx}
\usepackage{natbib}
\usepackage{color}

\def\del#1{{}}

\renewcommand{\phi}{\varphi}
\renewcommand{\vec}[1]{\bmath{#1}}
\newcommand{\vkp}{\vec{k}_\perp}
\newcommand{\kk}{\hat{k}}
\newcommand{\eps}{\epsilon}
\newcommand{\BB}{\overline{B}}
\newcommand{\vBB}{\overline{\vec{B}}}
\newcommand{\secref}[1]{Sec.~\ref{#1}}
\newcommand{\apref}[1]{Appendix~\ref{#1}}
\newcommand{\figref}[1]{Fig.~\ref{#1}}
\newcommand{\eqref}[1]{Eq.~(\ref{#1})}
\newcommand{\bea}{\begin{eqnarray}}
\newcommand{\eea}{\end{eqnarray}}
\renewcommand{\la}{\langle}
\newcommand{\ra}{\rangle}

\title[Probing magnetic turbulence by synchrotron polarimetry]
{Probing magnetic turbulence by synchrotron polarimetry:
  \\ statistics and structure of magnetic fields from Stokes correlators}
\author[A.\ H.\ Waelkens, A.\ A.\ Schekochihin and T.\ A.\ En{\ss}lin]
{A.\ H.\ Waelkens,$^{1}$\thanks{E-mail: waelkens@mpa-garching.mpg.de} 
A.\ A.\ Schekochihin$^{2,3}$ and T.\ A.\ En{\ss}lin$^{1}$ ,\\
$^1$Max-Planck-Institut f\"ur Astrophysik, Karl-Schwarzschild-Stra{\ss}e 1,
Postfach 1317, 85741 Garching, Germany\\
$^2$Rudolf Peierls Centre for Theoretical Physics, University of Oxford, 1 Keble Road, Oxford OX1 3NP, UK\\
$^3$Institut Henri Poincar\'e, Universit\'e Pierre et Marie Curie, 75231 Paris Cedex 5, France
}

\begin{document}
\pagerange{\pageref{firstpage}--\pageref{lastpage}}
\pubyear{2009}
\maketitle
\label{firstpage}

\begin{abstract}
We describe a novel technique for probing the statistical properties
of cosmic magnetic fields based on radio polarimetry data. Second-order 
magnetic field statistics like the power
spectrum cannot always distinguish between magnetic fields with essentially 
different spatial structure. Synchrotron polarimetry naturally allows certain 
4th-order magnetic field statistics to be inferred from observational data, 
which lifts this degeneracy 
and can thereby help us gain a better picture of the structure of the cosmic fields 
and test theoretical scenarios describing magnetic turbulence.
In this work we show that a 4th-order correlator of specific physical
interest, the tension-force spectrum, can be recovered from the polarized synchrotron 
emission data. 
We develop an estimator for this quantity based on polarized-emission observations 
in the Faraday-rotation-free frequency regime. We consider two cases: a
statistically isotropic field distribution, and a statistically
isotropic field superimposed on a weak mean field. In both cases the
tension force power spectrum is measurable; in the latter case,
the magnetic power spectrum may also be obtainable. 
The method is exact in the idealized case of a homogeneous relativistic-electron 
distribution that has a power-law energy spectrum with a spectral
index of $p=3$, and assumes statistical isotropy of the turbulent field.
We carry out numerical tests of our method using synthetic polarized-emission data 
generated from numerically simulated magnetic fields. 
We show that the method is valid, that it is not prohibitively sensitive 
to the value of the electron spectral index $p$, and
that the observed tension-force spectrum allows one to distinguish 
between, e.g., a randomly tangled magnetic field (a default assumption in many 
studies) and a field organized in folded flux sheets or filaments. 
\end{abstract}


\section{Introduction} \label{sect_intro}
Magnetized plasma is present almost everywhere in the observable
Universe, from stars and accretion disks to the interstellar
and the intracluster medium (respectively ISM and ICM). A large fraction of this magnetized
plasma is in a turbulent state. Understanding the origin of the cosmic 
magnetic fields and their evolution towards their observed state embedded 
in magnetized plasma turbulence, apart from being a tantalizing intellectual challenge 
in its own right \citep{Axel2005,Kandu2006,SC2005,SC2006,Schekochihin2009}, 
is also crucial in the construction of theories 
of large-scale dynamics and transport in many astrophysical systems. 
For example, magnetic fields are expected to be dynamically important in determining the 
angular momentum transport in accretion discs \citep{Pringle1972,Shakura1973}, 
to control star formation and the general structure of the ISM 
\citep[where magnetic fields prevent molecular clouds from
collapsing and suppress fragmentation; see, e.g.,][and references therein]{Price2008},
and to play an important role in galaxy discs as well as
galaxy clusters, where they influence the viscosity and thermal conductivity 
of the ISM and ICM \citep{Chandran1998,Medvedev2001,Markevitch2003} and 
the propagation of cosmic rays \citep[e.g.,][]{Strong2007,Yan2008}. 
Theoretical models of all these phenomena require some assumptions to be made 
about the spatial structure of the tangled magnetic fields permeating 
the constituent turbulent plasmas. However, as theory of magnetized plasma turbulence 
is in its infancy as a theoretical subject, there is no consensus about what 
this spatial structure is. In order to make progress both in understanding 
the turbulence and in modeling its effect on large-scale dynamics and transport, 
it is clearly desirable to be able to extract statistical information 
about the field structure from observational data \citep{EV2006,Ensslin2006}. 

\begin{figure*}
\begin{tabular}{c c} 
{\bf MHD } & {\bf Synthetic Gaussian} \\    
\includegraphics[width=0.45\textwidth]{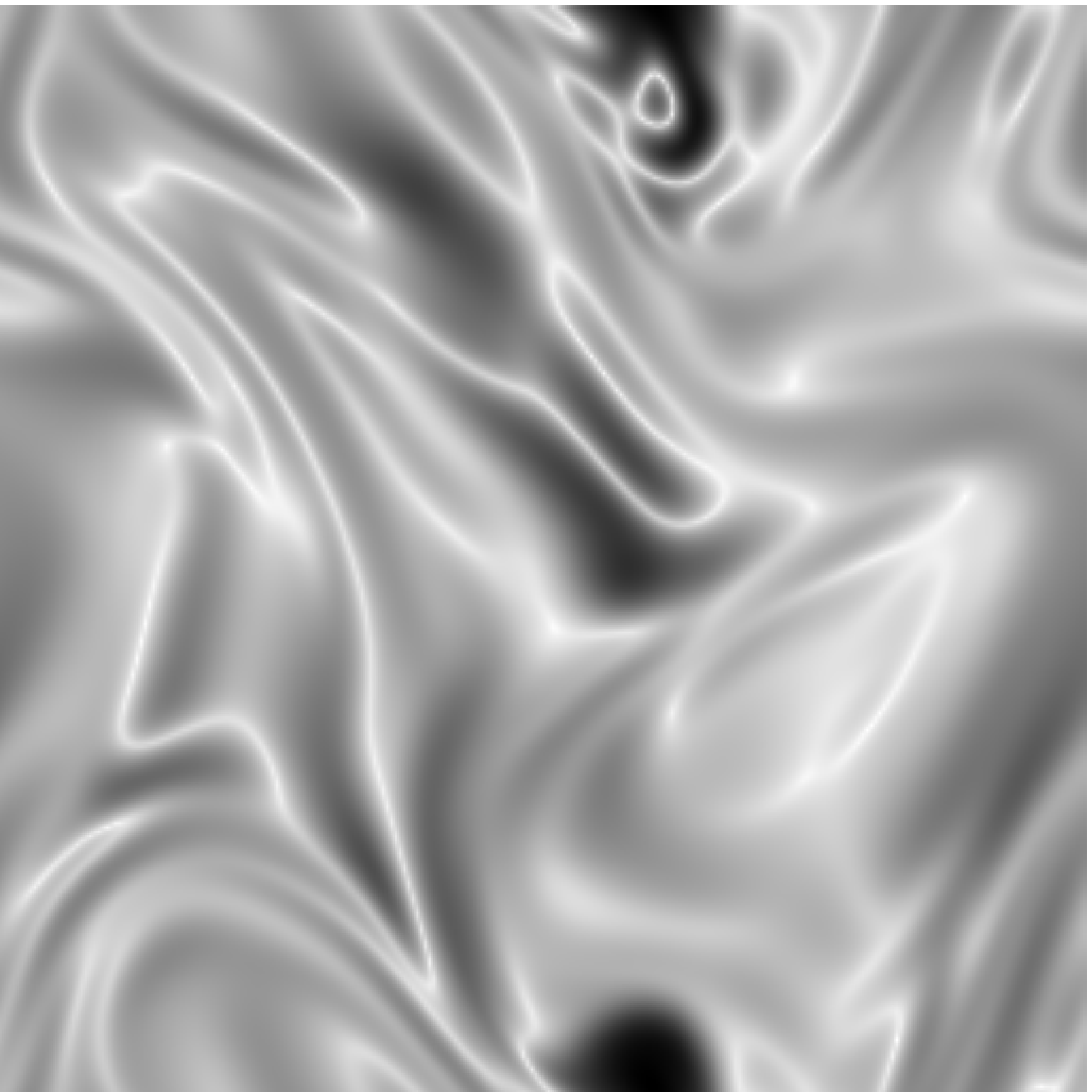} &
\includegraphics[width=0.45\textwidth]{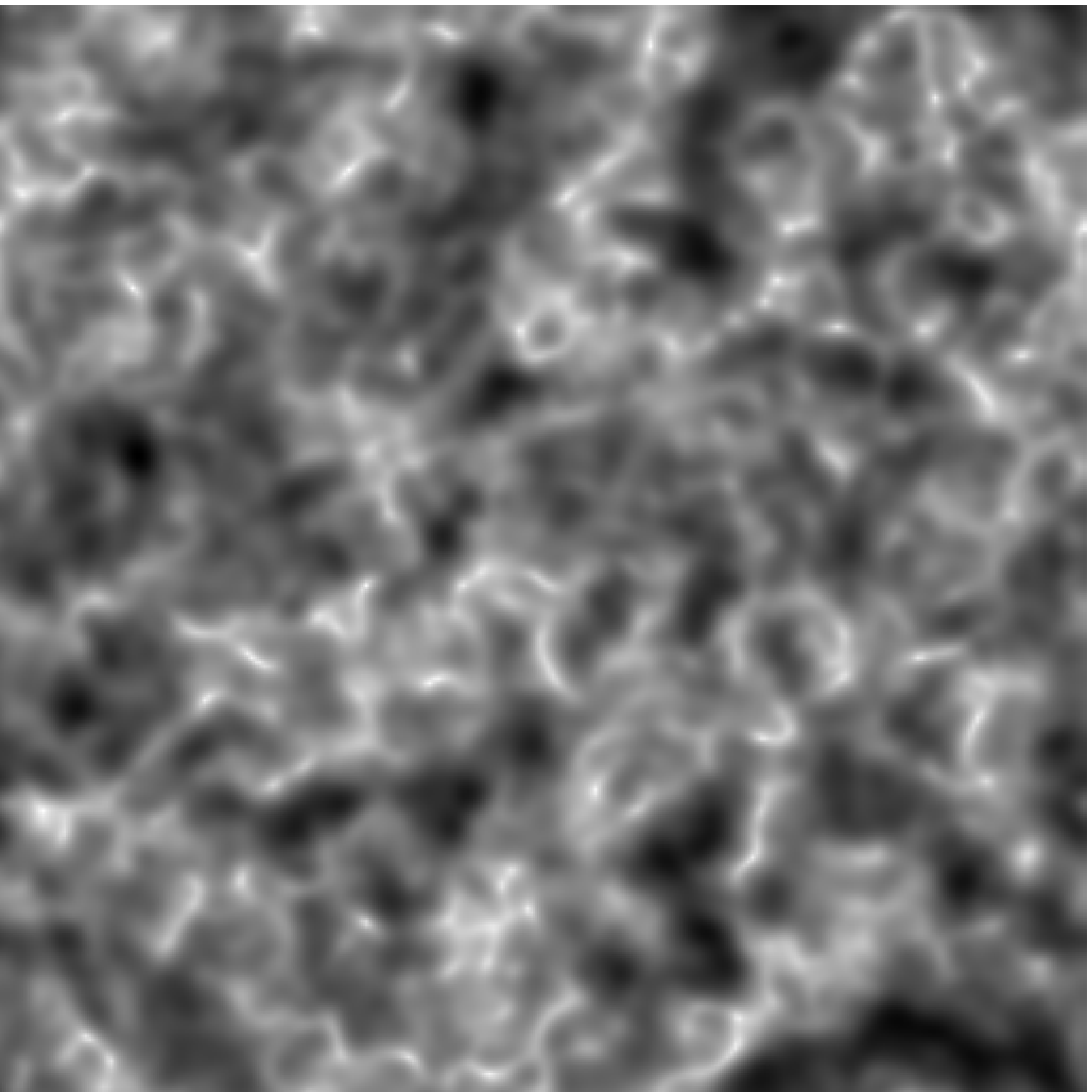} \\
\end{tabular}
\caption{{\it Left panel}: 
MHD-dynamo-generated magnetic field 
\citep{Alex2004} (saturated state of their Run S4).
{\it Right panel}: 
A synthetic divergence-free Gaussian random field realization with 
identical power spectrum. 
These are cross-sections of the field strength $|B|$ (dark represents
stronger field, white weaker field). 
The magnetic-field and tension-force power spectra are shown 
in \figref{fig::spectra}.}
\label{fig::TFpoint}
\end{figure*}

Diffuse synchrotron emission is observed throughout the
ISM and the ICM, as well as in the lobes of radio galaxies
\citep[e.g.][]{Westerhout1962,Wielebinski1962,Carilli1994,Reich2001,Beck2002,Wolleben2006,Haverkorn2006b,Reich2006,Clarke2006,Schnitzeler2007,Laing2008}. 
The fact that synchrotron emission is readily observable and is a good tracer of
the magnetic-field strength and orientation makes it a
key source of information that can serve as a reality check 
for theories of magnetized plasma turbulence and
magnetogenesis (origin of the magnetic fields). 

In this work, we will be focusing on how the 
synchrotron-emission data can be used to characterize the structure 
of the tangled magnetic fields permeating the ISM and the ICM. 
In this context we refer to previous studies which sought to recover 
statistical information about the structure of of these fields 
in the ICM from the Faraday rotation
measure (RM) data \citep{EV2003,Vogt2003,Vogt2005,Govoni2006,Guidetti2008}, as
well as studies of the ISM \citep{Haverkorn2006a,Haverkorn2008}, also
based on the RM data, and the work of \citet{Spangler1982,Spangler1983} 
and \citet{Eilek1989a,Eilek1989b} based on polarized synchrotron emission data. 
In formal terms, all of these papers are concerned with at most second-order statistics, 
namely the magnetic-field power spectrum, or the two-point correlation function of 
the magnetic field. 
Our work complements those previous efforts by 
drawing on the fact that polarized-emission data carries information 
about 4th-order statistics of the magnetic field.  
In particular, we present a practical method for
obtaining the tension-force power spectrum. 
As will be shown in greater detail in the following,
this quantity contains statistical information about 
the spatial structure of the tangled magnetic fields that is missing in the 
second-order statistics and, most importantly, 
is actually observable with radio telescopes mapping polarized synchrotron emission.

The plan of this paper is as follows.  In \secref{sec::why}, we
explain why the tension-force power spectrum is an interesting
quantity to measure and how it allows one to diagnose the
magnetic-field structure. In \secref{sec::method}, we explain  
the assumptions we make about the magnetic field 
(\secref{sec::mfield} and \secref{sec::assumptions}) 
and the observational data (\secref{sec::SP}; see also \apref{ap::SynRad}) 
and propose a method of reconstructing the tension-force power spectrum 
from the Stokes maps (\secref{sec::SC} and \secref{sec::TenForce}). 
We then generalize our method slightly for the case when a 
weak mean field is present and show that in this case 
the power spectrum of the magnetic field itself may be obtainable from the Stokes
correlators (\secref{sec::weakmf}). 
Most detailed analytical calculations required in this section 
are exiled to \apref{ap::theory}.
In \secref{sec::ASIM} we demonstrate the validity of our
method by testing it on synthetic observational data generated from
numerical simulations. A brief summary and conclusion is given in \secref{sec::C}.

\begin{figure*}
\centering
\includegraphics[angle=+90, width=0.45\textwidth]{./figs/M_z479_zsh260.ps}
\includegraphics[angle=+90, width=0.45\textwidth]{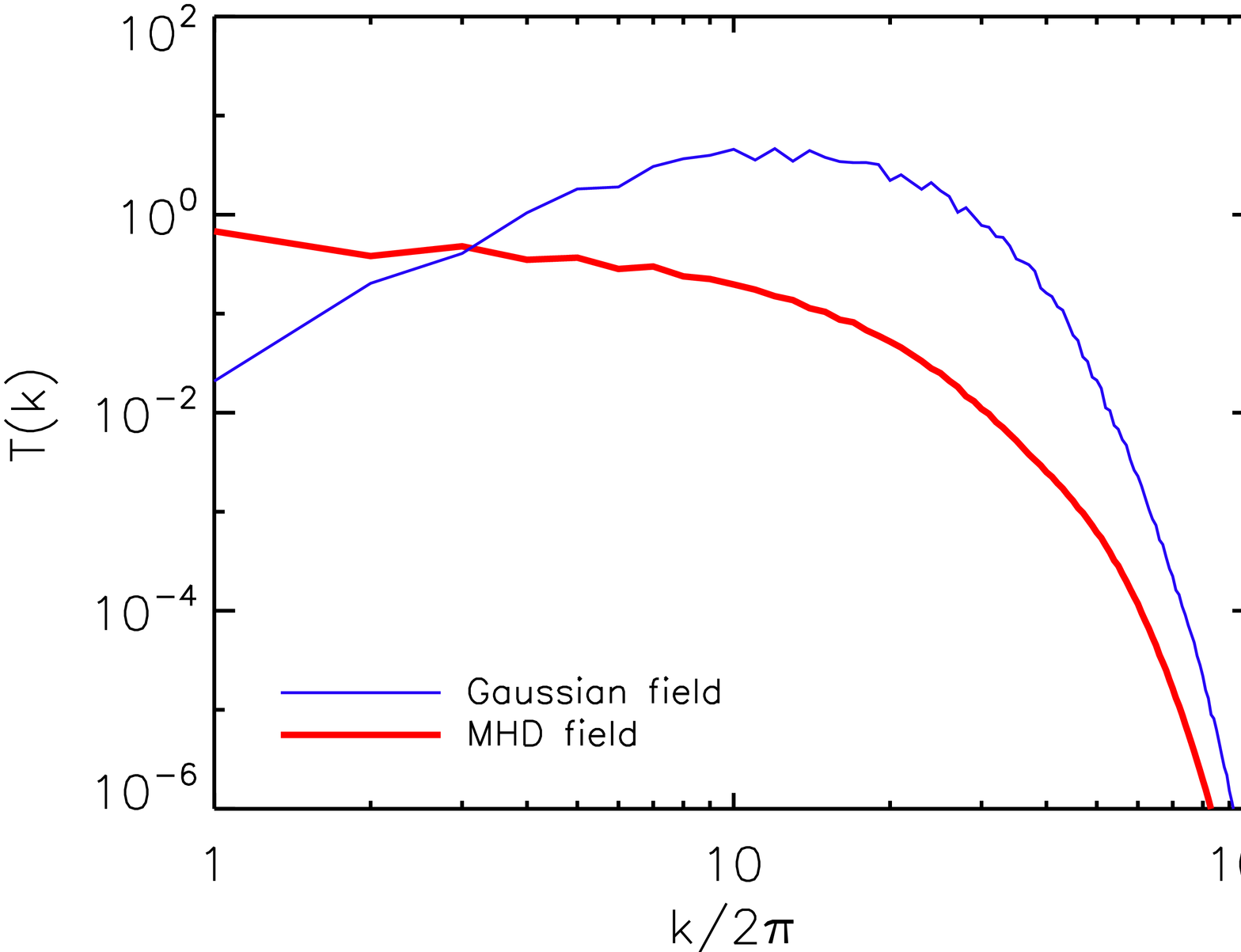}
\caption{{\it Left panel}: Magnetic-field power spectra for the fields shown in 
\figref{fig::TFpoint}. 
{\it Right panel}: Tension-force power spectra for the same fields.}
\label{fig::spectra}
\end{figure*}

\section{Motivation}
\label{sec::why}
Turbulent plasmas exhibit in general very complex magnetic structures (see 
\figref{fig::TFpoint}, left panel), which are best characterized by statistical means. 
The most widely used quantity for this purpose is the power spectrum 
\bea
M(k) =  4\pi k^2\la |\vec{B}(\vec{k})|^2 \ra,
\label{eq::PS}
\eea
where $\vec{B}(\vec{k})$ is the Fourier transform of the magnetic field 
(see \secref{sec::mfield}).
The angle-bracket averaging includes averaging over all directions of $\vec{k}$, 
so the power spectrum measures the amount of magnetic energy per wavenumber shell 
$|\vec{k}|=k$. It is related via the Fourier transform to the second-order two-point 
correlation function (or structure function) of the magnetic field.
It is an attractive quantity to measure because phenomenological theories of 
turbulence typically produce predictions for characteristic field increments 
between two points separated by a distance in the form of power laws with respect 
to that distance \citep{K41,Iroshnikov1963,Kraichnan1965,GS1995,Boldyrev2006,Schekochihin2009}---and 
such predictions are most obviously tested by 
measuring the spectral index (or the scaling exponent of the structure function). 
However, knowing the spectrum is not enough and can, in fact, be very misleading, 
for reasons having to do both with the physics of magnetic turbulence 
and with formal aspects of describing it quantitatively --- the basic point, 
which is discussed in much detail below, being that spectra do not contain 
any information about the geometrical structure of the magnetic-field lines. 

All scaling predictions for magnetized plasma turbulence 
proposed so far are, implicitly or explicitly, 
based on the assumption that magnetic fluctuations at sufficiently small scales
will look like small Alfv\'enic perturbations of a larger-scale ``mean'' field
(this is known as the \citealt{Kraichnan1965} hypothesis). 
Numerical simulations of MHD turbulence 
carried out without {\rm imposing} such a mean field do not appear to support 
this hypothesis \citep{Alex2004}, although the currently achievable resolution 
is not sufficient to state this beyond reasonable doubt and the results are to some 
extent open to alternative interpretations \citep{Haugen2004,Kandu2006}. 
What seems to be clear is that the magnetic field has a tendency to organize 
itself in long filamentary structures (``folds'') with field-direction 
reversals on very small scales \citep{Alex2004,Axel2005}. 
Filamentary magnetic structures are, indeed, observed in galaxy clusters 
\citep{Eilek2002,Clarke2006}, although 
the field reversal scale does not appear to be nearly as small as implied 
by MHD turbulence simulations---a theoretical puzzle solving which will probably 
require bringing in kinetic physics \citep[see discussion in][]{SC2006}. 

It is clear that both the current and future theoretical debates on 
the structure of magnetic turbulence would benefit greatly from 
being constrained observationally in a rigorous way. 
For the reasons explained above, in order to do this, we must be able to 
diagnose nontrivial spatial structure, which cannot be done by looking 
at the magnetic power spectrum alone. Let us explain this in more detail. 

Consider a divergence-free, helicity-free, statistically 
homogeneous and isotropic field as a minimal model for 
the fluctuating component of the magnetic field in galaxies and clusters. 
If this field also obeyed Gaussian statistics
exactly or, at least, approximately, its power spectrum
would be sufficient to completely describe its statistical properties
because all higher-order multi-point statistics 
could be expressed in terms of the second-order two-point correlators 
and, therefore, the power spectrum. 
Assuming such Gaussian statistics, \cite{Spangler1982, Spangler1983} and 
\cite{Eilek1989a} proposed to calculate the magnetic power spectrum 
using the observed total and polarized synchrotron 
radiation intensity, quantified by the Stokes parameters $I$, $Q$ and $U$
(see \apref{ap::SynRad}). 
Computing two-point correlation functions of the Stokes parameters, 
henceforth referred to as Stokes correlators, one essentially obtains 
two-point, 4th-order correlation functions of the magnetic field 
in the plane perpendicular to the line of sight (\secref{sec::SC}). 
If the statistics are Gaussian or if Gaussianity is adopted as a 
closure assumption, the 4th-order correlators can be split 
into second-order correlators, so the power spectrum follows. 

The problem with this approach to magnetic turbulence is that the Gaussian 
closure essentially assumes a structureless random-phased magnetic field, 
which then is, indeed, fully characterized by its power spectrum. 
It is evident in \eqref{eq::PS} that all phase information, which 
could tell us about the field structure, is lost in the power spectrum. 
As we explained above, both numerical and observational evidence 
(and, indeed, intuitive reasoning; see \citealt{Alex2004,SC2005}) 
show that magnetic fields do have structure and are very far from being 
a collection of Gaussian random-phased waves. Their spectra tell us little 
about this structure. This rather simple point is illustrated in 
\figref{fig::TFpoint}: the right panel depicts an instantaneous 
cross section of a 3D magnetic field obtained in a typical MHD dynamo 
simulation taken from \citet{Alex2004}, while the left panel 
shows a synthetically generated divergence-free Gaussian random field 
with exactly the same power spectrum (shown in \figref{fig::spectra}, left panel). 
The folded structure discussed above is manifest in the simulated field 
but absent in the Gaussian one: in the former case, the field typically varies
across itself on a much shorter scale than along itself and the
regions of strongest bending are well localized, whereas in the latter 
case, the field is uniformly tangled and has similar variation along and 
across itself. 

So how can one differentiate between such different fields in a systematic 
and quantitative way (i.e., other than by simply looking at visualizations)? 
As was pointed out by \citet{Schekochihin2002,Alex2004}, this can be done 
by looking at the statistics of the tension force 
\bea
\label{eq::TF}
\vec{F} = {\vec{B}\cdot\vec{\nabla}\vec{B}\over 4\pi}.
\eea
As a formal diagnostic, the tension force is a measure not just of the 
field strength but also of the gradient of the field 
along itself, thus it is strong if a field line is curved, and weak if the
field line is mostly straight. The tension-force field associated with a folded 
magnetic field (strong, straight direction-alternating fields in the ``folds'', 
weak curved fields in the ``bends'') will obviously be very different 
from the one associated with a random Gaussian field. 
As shown in \figref{fig::spectra} (right panel), their power spectra 
\bea
\label{eq::TFPS}
T(k) = 4\pi k^2 \la |\vec{F}(\vec{k})|^2 \ra
\eea
do, indeed, turn out to be very different: 
flat for the folded field, peaked at the smallest scales for the Gaussian field. 
Why a flat tension-force spectrum is expected for a folded field is discussed 
in \citealt{Alex2004}, their \S\,3.2.2, where numerical measurements of 
the tension-force statistics can also be found. In contrast, 
for the Gaussian field, one obviously gets $T(k)\propto k^2M^2$, 
hence the peak at the small scales. 

In physical terms, 
the tension force is one of the two components of the Lorentz force
\begin{eqnarray}
\frac{1}{c}\,\vec{J}\times\vec{B}=-\vec{\nabla}\frac{B^2}{8\pi}
+{\vec{B}\cdot\vec{\nabla}\vec{B}\over4\pi},
\end{eqnarray}
where the first term on the right-hand side is the
magnetic pressure force, and the second term is the magnetic tension
force, as defined in \eqref{eq::TF}. In subsonic turbulence, 
the tension force essentially determines the dynamical back reaction 
of the magnetic field on the plasma motions because regions with
higher magnetic pressure can be expected to have correspondingly weaker
thermal pressure, so that the magnetic pressure
forces are mostly balanced by oppositely directed thermal pressure forces. 

Thus, measuring tension-force power spectra not only permits one to
discriminate quantitatively between 
different magnetic turbulence scenarios but 
also provides a detailed insight into the MHD physics occurring in space, because it
quantifies the properties of the dynamically relevant force in the
magnetic turbulence. It is perhaps worth stressing this last point. 
In principle, many 4th-order statistical quantities 
that one might construct out of the Stokes correlators should
be able to discern between different magnetic-field structures, but 
the tension-force power spectrum also has a clear physical meaning. 

It is a stroke of luck that not only the tension-force power spectrum 
is the diagnostic that we would ideally like to know from the theoretical 
point of view, but it turns out that, under mild simplifying assumptions, 
it can be fully recovered from the statistical information contained  
in the Stokes correlators and, therefore, it is observable! 
This will be demonstrated in detail in the following 
sections. Such an outcome is not automatic: other potentially interesting statistical 
quantities such as the magnetic-energy power spectrum or the magnetic pressure-force 
statistics are not so directly imprinted into the Stokes correlators
and require further assumptions in order to be
extractable from the same data. 

\section{Method}
\label{sec::method}

In this section, we outline a formal theoretical framework 
for converting polarized-emission observables into the physically 
interesting statistical characteristics of the magnetic field 
under a number of simplifying assumptions. 

\subsection{Magnetic Field}
\label{sec::mfield}

Let us assume some volume $V$ of interstellar or intracluster plasma 
to be filled with a magnetic field $\vec{B}(\vec{x})$ 
and a magnetized relativistic electron population giving rise to the 
synchrotron emission we observe (\figref{fig::S}).
We use a Cartesian coordinate system $(x,y,z)$, where $z$ is the line of sight. 
The volume under consideration is assumed to have depth $L$ in this direction. 
The magnetic field can be decomposed into two parts:
\bea
\vec{B}=\vBB + \vec{b} ,
\label{eq::Bsubdiv}
\eea
where $\vBB = \la\vec{B}\ra$ is the regular (mean) field throughout the volume
under consideration and $\vec{b}$ is the fluctuating (``turbulent'') field.
The former is assumed to be known and the latter is what we aim to study. 
We will work out its various correlation functions and their relationship 
to observable quantities---this can be done both in 
position space and in Fourier space in largely analogous ways. 
The Fourier transform of the field is defined according to 
\bea
\hat{\vec{b}}(\vec{k}) = {1\over V}\int d^3\vec{x}\, e^{-i\vec{k}\cdot\vec{x}}\vec{b}(\vec{x}) ,
\qquad
\vec{b}(\vec{x}) = \sum_{\vec{k}} e^{i\vec{k}\cdot\vec{x}}\hat{\vec{b}}(\vec{k}) ,
\label{eq::FTdef}
\eea
where $\vec{x}=(x,y,z)$. 
In what follows we will drop the hats on the Fourier transformed quantities. 
Note that discrete and continuous wave-vector spaces are related via a simple 
mnemonic: 
\bea
\sum_{\vec{k}} \Leftrightarrow {V\over(2\pi)^3}\int d^3\vec{k}.
\eea

\subsection{Assumptions: Homogeneity and Isotropy}
\label{sec::assumptions}

We will make two key assumption about the fluctuating magnetic field: 
statistical homogeneity and isotropy. The first of these is not a serious restriction 
of generality as, essentially, we would like to calculate statistical information 
based on data from subvolumes within which system-size spatial variation 
of the bulk properties of the astrophysical plasma under consideration 
can be ignored. The second assumption, the isotropy, is more problematic 
because of the known property of magnetized turbulence to be strongly 
anisotropic with respect to the direction of the mean field, {\em provided 
the mean field is dynamically strong} \citep[see discussion and exhaustive 
reference lists in][]{SC2005,Schekochihin2009}. 
It will, therefore, only be sensible to apply our method to astrophysical
situations where the mean field is either absent or weak, 
i.e., $\BB^2  \ll \la |\vec{b}|^2 \ra$. This should be a very good 
approximation for the ICM and may also be reasonable in parts of the ISM 
\citep[e.g., in the spiral arms; see][]{Haverkorn2006a,Haverkorn2008}. 

In what follows, 
we will first consider the case of $\BB=0$ and then provide 
a generalization of our results to the case of a weak mean field 
(\secref{sec::weakmf}). In both cases, we will first show how far 
one can get without the isotropy assumption and then find that 
only assuming isotropy are we able to calculate 
the tension-force power spectrum. It will also turn 
out that, in the case of a non-zero weak mean field, 
additional information can be gleaned from polarized-emission data, including 
the power spectrum of the fluctuating field (normally not available without 
the Gaussian closure, as discussed in \secref{sec::why}). 

Physically, we might argue that a weak mean field does not modify
the turbulent dynamics and, therefore, does not break the statistical 
isotropy of the small-scale turbulent field. Obviously, if the bulk of 
the magnetic energy turns out to reside above or at some characteristic scale $l_B$, 
the statistically isotropic fluctuating field at that scale will look like 
a (strong) mean field to fluctuations at scales smaller than $l_B$ and 
assuming isotropy of those fluctuations will almost certainly be wrong. 
Thus, our method can only be expected to handle successfully magnetic fluctuations 
at scales larger than $l_B$. This, however, is sufficient to make the outcome 
interesting because the key question in the theoretical discussions 
about the nature of the cosmic magnetic turbulence referred to in \secref{sec::why} 
is precisely what determines $l_B$ (is it the reversal scale of the folded fields? 
what is that scale?) and how diagnosing the spatial structure 
of the field at scales above $l_B$ might help us answer this question. 

Note that a field organized in folds or filaments, as in \figref{fig::TFpoint} (left panel), 
is statistically isotropic because, while the folds extend over long distances, 
their orientation is random. 

\subsection{Observables: Stokers Parameters}
\label{sec::SP}

Our direct observable is the partially linearly polarized synchrotron emission of 
the relativistic electrons gyrating in the magnetic field. This emission 
is measured by radio telescopes in projection onto the sky 
in terms of the Stokes parameters $I$, $Q$ and $U$. Let us briefly recapitulate 
the relevant physics. 

We assume a relativistic electron population that is spatially homogeneous, 
has an isotropic pitch-angle distribution, and  
a power-law energy distribution: 
\bea
\label{eq::eldistr}
N(\gamma) d\gamma = C \gamma^{-p} d\gamma,
\eea 
where $\gamma$ is the Lorentz factor, $N$ the number of 
electrons per $\gamma$ per unit volume and $C$ is the normalization 
factor proportional to the electron density. 
The observed emission will then be partially
linearly polarized \citep{RL} and, therefore, 
at any given observed (radio) frequency $\nu$, it is fully characterized by
the Stokes parameters $I$, $Q$ and $U$ as functions of the sky coordinate
$\vec{x}_\perp=(x,y)$ 
(the spatial coordinate in the plane perpendicular to the line of sight). 
This is detailed in \apref{ap::SynRad}. 

As explained in \apref{ap::SynRad}, a measurement of the absolute values of 
$I$, $Q$ and $U$ (and, therefore, of the magnetic field and its tension)
requires knowledge of the relativistic-electron energy density, which 
enters via the factor $C$ in \eqref{eq::eldistr}. In our Galaxy, it can be 
inferred directly from its value measured at Earth, or via independent 
messengers such as inverse Compton emission of starlight and cosmic-microwave-background 
photons. The absolute rms value of the magnetic field can then be inferred 
either via the Stokes parameters or via the popular assumption of 
equipartition between the average magnetic energy density and the 
relativistic-electron energy
density \citep[see, e.g.][and references therein]{Beck2005}. 
Note that it is essential for the method developed below that the relativistic 
electrons can be assumed to have a homogeneous distribution ($C={\rm const}$), 
and do not follow the magnetic-field-strength enhancements on small scales. 
Otherwise our method would provide incorrect results, since such a local coupling
of relativistic electrons and magnetic fields is not incorporated, and would destroy the
assumed relation between the spatial variation of the synchrotron observables 
and the magnetic fields. However, since the relativistic electrons are very diffusive 
along and even perpendicular to the magnetic fields, a roughly homogeneous distribution 
can safely assumed in most relevant environments.

We further assume $p=3$ in \eqref{eq::eldistr} 
(corresponding to the frequency distribution $\propto\nu^{-1}$). 
This is a convenient choice because then all Stokes parameters 
are quadratic in the magnetic field, which means that their two-point 
correlation functions will give us 4th-order statistics. 
We stress that this power law, although expected by theoretical shock acceleration models
\citep{Drury1983}, is, of course, a simplification of reality \citep[see, e.g.][]{Strong2007}. 
However, it is usually a sufficiently good approximation over fairly wide frequency ranges 
for many synchrotron sources. Thus, $p=3$ is reasonably close to the values observed for 
our own Galaxy \citep{Reich1988,Tateyama1986}, the values obtained by
CMB foreground subtraction techniques \citep[e.g.][]{Tegmark1996,Angelica2008,Jo2008,Bottino2008}, 
and found in extra-galactic observations of radio-galaxies \citep{Beck1996}. 
While the theoretical developments that follow do depend on taking $p=3$, 
the numerical tests of the resulting method reported in \secref{sec::spec_ind} 
will show that it is not essential that $p=3$ be satisfied particularly precisely. 
Deviations from $p=3$ can be addressed analytically in a more quantitative way by a
Taylor expansion around $p=3$, which we leave for further work. 

Finally, we assume the observed volume to be optically
thin and its Faraday depth to be negligible at the observation
frequency $\nu$. At high frequencies, both conditions tend to be
satisfied, Faraday rotation being in most cases the greater
constraint. For example, in our Galaxy, Faraday rotation is a
relevant phenomenon at frequencies below a few GHz, while the medium remains
mostly optically thin down to frequencies of a few hundred MHz, where
free-free absorption starts being relevant \citep[see][and references therein]{Sun2008}. 
In cases where Faraday rotation is present in the
frequency range of the data, we assume that a Faraday de-rotation has been 
applied. Even in the case of source-intrinsic Faraday rotation, this can still
be achieved using Faraday tomography techniques \citep{Brentjens2005}. 

Under these conditions the Stokes parameters can be written as 
the following line-of-sight integrals
\bea
I(\vec{x}_\perp) & = & {1\over L}\int_0^L dz 
\left[ B_x^2(\vec{x}) + B_y^2(\vec{x})\right],\nonumber\\
Q(\vec{x}_\perp) & = & {1\over L}\int_0^L dz 
\left[ B_x^2(\vec{x}) - B_y^2(\vec{x})\right], \nonumber\\
U(\vec{x}_\perp) & = & {1\over L}\int_0^L dz \,2B_x(\vec{x})B_y(\vec{x}),
\label{eq::StoPar}
\eea 
where $L$ is the depth of the emission region. 
The dimensional prefactors converting the magnetic-field strength to radio 
emissivity have been suppressed (see \apref{ap::SynRad}). 

\begin{figure}
  \centering
  \includegraphics[width=0.3\textwidth]{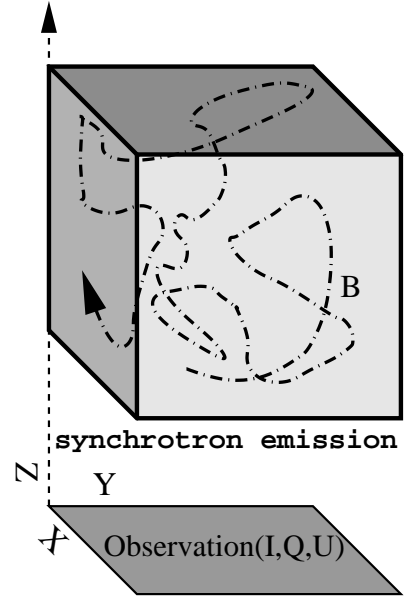}
  \caption{Magnetic field and the observables: artist's impression.}
  \label{fig::S}
\end{figure}

\subsection{From Stokes Correlators to Magnetic-Field Statistics}
\label{sec::SC}

Thus, observed polarized emission provides us with three scalar fields related 
quadratically to the magnetic field projected onto the plane perpendicular 
to the line of sight. We can construct 6 two-point correlators of these 
fields, which we will refer to as the Stokes correlators: 
\bea
\label{eq::SCdef}
\Sigma_{XY}(\vec{r}_\perp)=\langle
X(\vec{x}_\perp) \, Y(\vec{x}_\perp+\vec{r}_\perp)\rangle,
\eea
where $X,Y \in \{ I,Q,U\}$ and $\langle ... \rangle$ denote a statistical average 
performed over the observational maps, which usually means volume averaging with 
respect to the sky coordinate $\vec{x}_\perp$.

Are the Stokes correlators sufficient to reconstruct the statistics 
of the magnetic field? 

In formal terms, the statistical properties of a stochastic field 
are fully described by its $n$-point distribution function, or, 
equivalently, by the full set of its $n$-point, $m$-th order 
correlation tensors. In practice, this is too much information, 
most of it is not observable in any realistic situation, 
and in any event, only a few of these correlators can be interpreted in 
simple physical terms and are, therefore, useful for a qualitative 
understanding of the field structure. As the Stokes correlators 
are 4th order in the magnetic field and measure its correlations between 
two points in space, it is the two-point, 4th-order correlation tensor 
that will be relevant to this discussion: 
\bea
\nonumber
C_{ij,mn}(\vec{r}) 
&=& \la B_i(\vec{x})B_j(\vec{x})B_m(\vec{x}+\vec{r})B_n(\vec{x}+\vec{r})\ra\\
&=& \la H_{ij}(\vec{x})H_{mn}(\vec{x}+\vec{r})\ra,
\label{eq::Cijmn}
\eea
where, for notational convenience, we have introduced the field tensor $H_{ij}=B_iB_j$. 
The angle brackets denote statistical average, understood ideally as an 
ensemble (or time) average and in practice, if we are dealing with one 
observed realization of the field, as the volume average: 
$\la\dots\ra = (1/V)\int d^3\vec{x}(\dots)$. 
Implicitly, performing a volume average relies on the assumption 
of statistical homogeneity (\secref{sec::assumptions}), i.e., independence 
of the statistical properties of the field of the reference 
point $\vec{x}$ where they are calculated. 
In terms of Fourier-space quantities, we have 
\bea
\label{eq::Ck}
C_{ij,mn}(\vec{r}) = \sum_{\vec{k}} e^{i\vec{k}\cdot\vec{r}} C_{ij,mn}(\vec{k}),
\quad C_{ij,mn}(\vec{k}) = \la H_{ij}^*(\vec{k}) H_{mn}(\vec{k})\ra,
\eea
where the Fourier transforms of all quantities are defined similarly to \eqref{eq::FTdef}. 

In general, the tensor $C_{ij,mn}$ depends on very many 
independent scalar functions, so the 6 available 
Stokes correlators [\eqref{eq::SCdef}] cannot 
provide all the required information necessary to recover the magnetic-field statistics. 
Indeed, let us write the Stokes correlators 
in terms of the correlation tensor $C_{ij,mn}$. It is particularly easy to 
do this in Fourier space because the line-of-sight integration in 
\eqref{eq::StoPar} amounts simply to picking the $k_z=0$ component 
of the field:
\bea
\nonumber
I(\vkp) &=& H_{xx}(\vkp) + H_{yy}(\vkp),\\ 
\nonumber
Q(\vkp) &=& H_{xx}(\vkp) - H_{yy}(\vkp),\\ 
U(\vkp) &=& 2H_{xy}(\vkp), 
\eea 
where $\vkp=(k_x,k_y,0)$. 
Therefore, the Fourier transforms of the Stokes correlators 
[\eqref{eq::SCdef}] are 
\bea
\Sigma_{II}(\vkp) & = & 
C_{xx,xx}(\vkp) + C_{xx,yy}(\vkp) + C_{xx,yy}^*(\vkp) + C_{yy,yy}(\vkp), \nonumber\\
\Sigma_{QQ}(\vkp) & = & 
C_{xx,xx}(\vkp) - C_{xx,yy}(\vkp) - C_{xx,yy}^*(\vkp) + C_{yy,yy}(\vkp), \nonumber\\
\Sigma_{UU}(\vkp) & = & 4 C_{xy,xy}(\vkp), \nonumber\\
\Sigma_{IQ}(\vkp) & = & 
C_{xx,xx}(\vkp) - C_{xx,yy}(\vkp) + C_{xx,yy}^*(\vkp) - C_{yy,yy}(\vkp), \nonumber\\
\Sigma_{IU}(\vkp) & = & 
2 \left[C_{xx,xy}(\vkp) + C_{yy,xy}(\vkp)\right], \nonumber\\
\Sigma_{QU}(\vkp) & = & 
2 \left[C_{xx,xy}(\vkp) - C_{yy,xy}(\vkp)\right]. 
\label{eq::SC}
\eea
This immediately implies that 
\bea
\nonumber
C_{xx,xx}(\vkp) &=& 
{1\over4}\left[\Sigma_{II}(\vkp) + \Sigma_{QQ}(\vkp)
+ 2{\rm Re}\,\Sigma_{IQ}(\vkp)\right] ,\\
\nonumber
C_{yy,yy}(\vkp) &=& 
{1\over4}\left[\Sigma_{II}(\vkp) + \Sigma_{QQ}(\vkp)
- 2{\rm Re}\,\Sigma_{IQ}(\vkp)\right] ,\\
\nonumber
C_{xx,yy}(\vkp) &=& 
{1\over4}\left[\Sigma_{II}(\vkp) - \Sigma_{QQ}(\vkp) 
-i 2{\rm Im}\,\Sigma_{IQ}(\vkp)\right],\\
\nonumber
C_{xy,xy}(\vkp) &=& {1\over4}\Sigma_{UU}(\vkp),\\
\nonumber
C_{xx,xy}(\vkp) &=&
{1\over4}\left[\Sigma_{IU}(\vkp) + \Sigma_{QU}(\vkp)\right],\\
C_{yy,xy}(\vkp) &=&
{1\over4}\left[\Sigma_{IU}(\vkp) - \Sigma_{QU}(\vkp)\right].
\label{eq::CxySC}
\eea
These are the only components of the correlation tensor \eqref{eq::Ck} that 
are observable directly and it is only their dependence on the wave vector 
perpendicular to the line of sight that can be probed. 
No correlators that involve the projection of the field on the line of sight 
($B_z$) can be known.

The number of independent scalar functions that determine $C_{ij,mn}$ 
is reduced and becomes closer to the number of observables if 
we make some symmetry assumptions and, in particular, isotropy (\secref{sec::assumptions}). 
Under this assumption, it turns out that we only need to know 
7 independent scalar functions of $k=|\vec{k}|$ to reconstruct $C_{ij,mn}$ fully 
(see \apref{ap::4thorder}). 
It also turns out that, if no mean field is present ($\BB=0$), 
only 4 of the Stokes correlators of an isotropic field 
contain independent information: 
$\Sigma_{II}$, two of $\Sigma_{QQ}$, $\Sigma_{UU}$, $\Sigma_{QU}$ 
and one of $\Sigma_{IQ}$, $\Sigma_{IU}$. For example, if we keep 
$\Sigma_{II}$, $\Sigma_{QQ}$, $\Sigma_{UU}$ and $\Sigma_{IQ}$, 
the other two Stokes correlators are 
\bea
\nonumber
\Sigma_{IU} &=& \Sigma_{IQ}\tan2\phi,\\
\Sigma_{QU} &=& {1\over2}\left(\Sigma_{QQ}-\Sigma_{UU}\right)\tan4\phi,
\label{eq::IU_QU_rlns}
\eea
where $\phi$ is the angle between $\vkp$ and the $x$ axis of the frame 
in which the Stokes parameters are measured, i.e., 
$\vkp = k_\perp(\cos\phi,\sin\phi, 0)$. 
These relations are useful in constructing well behaved expressions 
for the observables (see \secref{sec::TenForce} and \apref{ap::GenObs}). 
They could also be useful in practical situations when the Stokes 
maps might not be perfect, so one might have more (or higher-quality) 
data on some Stokes correlators than on others. 

We see that, even with isotropy, we do not have enough observables to 
measure the general 4th-order statistics of the magnetic field
(7 independent scalar functions needed, 4 available). 
However, the information carried by the Stokes correlators does 
suffice to reconstruct some of the correlation functions of the field. 
How to determine whether any particular 4th-order correlator is 
observable is explained in \apref{ap::GenObs}. In a stroke of luck, 
we find that we can reconstruct the tension-force power spectrum, 
which is a physically interesting quantity because it diagnoses 
the geometrical structure of the magnetic field and 
its dynamical action on the plasma motions (\secref{sec::why}). 
Although it follows from the general procedure given in \apref{ap::GenObs}
(see \apref{ap::TenForce}), it is perhaps illuminating to provide an individual 
derivation for this quantity. 

\subsection{Tension-Force Power Spectrum}
\label{sec::TenForce}

The tension force [\eqref{eq::TF}] is $F_i = B_j\partial_j B_i = \partial_j H_{ij}$,  
where we have omitted the factor of $1/4\pi$). Therefore, its spectrum [\eqref{eq::TFPS}] 
is 
\bea
\label{eq::Tdef}
T(k) = 4\pi k^2\Phi(k), 
\eea
where 
\bea
\label{eq::MTFT}
\Phi(k) = \la F_i^*(\vec{k})F_i(\vec{k})\ra = k_j k_n C_{ij,in}(\vec{k}). 
\eea
We do not have any directly observable information about $k_z\neq 0$, so let us 
set $\vec{k}=\vkp = k_\perp(\cos\phi,\sin\phi, 0)$.
Then 
\bea
\Phi(k_\perp) = \Phi_1 + \Phi_2,
\eea
where $\Phi_1$ is the part that is directly recoverable from the Stokes correlators 
[using \eqref{eq::CxySC}]:
\bea
\nonumber
\Phi_1 &=& k_x^2 \left[C_{xx,xx}(\vkp) + C_{xy,xy}(\vkp)\right] 
+ k_y^2 \left[C_{xy,xy}(\vkp) + C_{yy,yy}(\vkp)\right]\\ 
\nonumber
&&+\,\, 2k_xk_y {\rm Re}\left[C_{xx,xy}(\vkp) + C_{yy,xy}^*(\vkp)\right]\\
&=& {1\over 4}\,k^2 \left[\Sigma_{II} + \Sigma_{QQ} + \Sigma_{UU} + 
2{\rm Re}\left(\Sigma_{IQ}\cos2\phi + \Sigma_{IU}\sin2\phi\right)\right],
\nonumber\\
\label{eq::Phi1}
\eea
whereas $\Phi_2$ is the part that contains magnetic-field components parallel to the line of sight 
and, therefore, not picked up by the polarized-emission observations:
\bea
\label{eq::Phi2}
\Phi_2 = k_x^2 C_{xz,xz}(\vkp) + k_y^2C_{yz,yz}(\vkp) + 2k_xk_y {\rm Re}C_{xz,yz}(\vkp).
\eea
It is to reconstruct this missing information that we have to assume isotropy, 
because it gives us a symmetry relationship between the unobservable correlators 
and the observable ones. If no mean field is present ($\BB=0$), it is possible 
to show (see \apref{ap::TenForce}) that, for a statistically isotropic magnetic-field 
distribution,
\bea
\nonumber
\Phi_2 &=& k^2\left\{C_{xy,xy}(\vkp) - {k_x k_y\over k_x^2-k_y^2}
\left[C_{xx,xy}(\vkp) - C_{yy,xy}(\vkp)\right]\right\}\\
&=& {1\over4}\,k^2\left(\Sigma_{UU} - \Sigma_{QU}\tan 2\phi\right).
\label{eq::Phi2iso}
\eea

Assembling the directly observable [\eqref{eq::Phi1}] and the inferred 
[\eqref{eq::Phi2iso}] 
part of the tension-force power spectrum, we arrive at an expression for 
$\Phi(k)$ solely in terms of the Stokes correlators. 
There are two further steps that need to be taken to bring this expression 
into a practically computable form. 

Firstly, let us recall that, 
while the Stokes correlators in \eqref{eq::Phi1} and \eqref{eq::Phi2iso} 
depend on the vector $\vkp$, 
the tension-force spectrum $\Phi$ must depend only on $k=|\vkp|$. It is, therefore, 
permissible (and, in fact, increases the quality of the statistics) to average 
our expression for $\Phi$ over the angle $\phi$ (i.e., over 
a shell $|\vkp|=k$ in the wavenumber space). 

Secondly, the fact that, for an isotropic field, only 4 of the 6 
available Stokes correlators are independent [see \eqref{eq::IU_QU_rlns}] 
can be used to construct many theoretically equivalent expressions for $\Phi(k)$. 
Additional freedom comes from the angle independence of $\Phi(k)$ and, 
therefore, the possibility of doing weighted angle averages (see \apref{ap::GenObs}). 
The strategy for choosing a particular formula for practical computations is 
to avoid having singularities in the coefficients: such as the factor 
of $\tan2\phi$ in \eqref{eq::Phi2iso}. How to do this systematically 
is explained in \apref{ap::GenObs}, but here we simply give the 
result: 
\bea
\nonumber
T(k) \!\!\!\!\! &=& \!\!\!\!\!
{1\over2}\,k^4\int_0^{2\pi} d\phi\, \biggl[\Sigma_{II} 
+ 2\left(\Sigma_{IQ}\cos2\phi + \Sigma_{IU}\sin2\phi\right)
- \Sigma_{QU}\sin4\phi
\biggr.\\
&&+\,\,\biggl. {1\over2}\left(3-\cos4\phi\right)\Sigma_{QQ}
+ {1\over2}\left(3+\cos4\phi\right)\Sigma_{UU}\biggr].
\label{eq::Tk_obs}
\eea
This formula is derived in \apref{ap::TenForce} from our general method, 
but can also be easily seen to follow directly from \eqref{eq::Phi1} and 
\eqref{eq::Phi2iso} via \eqref{eq::IU_QU_rlns}, angle averaging 
and multiplication by the wave-number-space volume factor of $4\pi k^2$ 
[see definition of $T(k)$, \eqref{eq::Tdef}]. 
\eqref{eq::Tk_obs} is our final expression 
for the tension-force power spectrum. 

Thus, we have accomplished our goal of showing that, 
despite the scarcity of the observable information, 
the tension-force power spectrum can be fully reconstructed 
from the available Stokes correlators (in \apref{ap::GenObs}, we 
also show how to construct all other observable 4th-order quantities). 
In \secref{sec::ASIM}, we will test our method of doing this, but 
first, we generalize it slightly to the case of weak mean field. 

\subsection{Case of Weak Mean Field: Observing the 
Magnetic-Field Power Spectrum}
\label{sec::weakmf}

We now relax the assumption that $\vBB=0$ in \eqref{eq::Bsubdiv}. 
Then the 4th-order correlation tensor $C_{ij,mn}$ [\eqref{eq::Cijmn}] 
can be written in terms of the mean field and of the correlation tensors of 
the fluctuating field: 
\bea
\nonumber
C_{ij,mn}(\vec{r}) &=& \BB_i\BB_j\BB_m\BB_n 
+ \BB_i\BB_j \la b_m' b_n'\ra + \BB_m\BB_n \la b_i b_j\ra\\
&& +\,\, \BB_i\BB_m \la b_j b_n'\ra + \BB_i\BB_n \la b_j b_m'\ra
\nonumber\\
&& +\,\, \BB_j\BB_m \la b_i b_n'\ra + \BB_j\BB_n \la b_j b_m'\ra
\nonumber\\
&& +\,\, \BB_i \la b_j b_m' b_n'\ra + \BB_j \la b_i b_m' b_n'\ra
\nonumber\\
&& +\,\, \BB_m \la b_i b_j b_n'\ra + \BB_n \la b_i b_j b_m'\ra
\nonumber\\
&& +\,\, \la b_i b_j b_m' b_n'\ra,
\label{eq::Cgeneral}
\eea
where unprimed quantities are evaluated at $\vec{x}$ and the primed ones 
at $\vec{x}+\vec{r}$. Due to homogeneity, correlation tensors depend 
only on $\vec{r}$ and not on $\vec{x}$ (and the statistical average can 
be interpreted as a volume average over $\vec{x}$). This means that 
the first three terms in \eqref{eq::Cgeneral} have no spatial dependence 
at all, while the rest of the tensor can be written in Fourier space 
as follows:
\bea
\nonumber
C_{ij,mn}(\vec{k}) &=& {1\over V}\int d^3\vec{r}\, e^{-i\vec{k}\cdot\vec{r}} C_{ij,mn}(\vec{r})\\
&=& \BB_i\BB_m c_{j,n}(\vec{k}) + \BB_i\BB_n c_{j,m}(\vec{k})
\nonumber\\
&& +\,\, \BB_j\BB_m c_{i,n}(\vec{k}) + \BB_j\BB_n c_{i,m}(\vec{k})
\nonumber\\
&& +\,\, \BB_i c_{mn,j}^*(\vec{k}) + \BB_j c_{mn,i}^*(\vec{k})
\nonumber\\
&& +\,\, \BB_m c_{ij,n}(\vec{k}) + \BB_n c_{ij,m}(\vec{k})
\nonumber\\
&& +\,\, c_{ij,mn}(\vec{k}).
\label{eq::Ckgeneral}
\eea
This is the Fourier-space correlation tensor introduced in \eqref{eq::Ck}, 
which has now been expressed in terms of the mean field and 
the second-, 3rd- and 4th-order correlation tensors of the 
fluctuating field:
\bea
c_{i,m}(\vec{k}) &=& \la b_i^*(\vec{k}) b_m(\vec{k})\ra 
= {1\over V}\int d^3\vec{r}\, e^{-i\vec{k}\cdot\vec{r}} 
\la b_i b_m'\ra,\\
c_{ij,m}(\vec{k}) &=& \la h_{ij}^*(\vec{k}) b_m(\vec{k})\ra
= {1\over V}\int d^3\vec{r}\, e^{-i\vec{k}\cdot\vec{r}} 
\la b_i b_j b_m'\ra,\\
c_{ij,mn}(\vec{k}) &=& \la h_{ij}^*(\vec{k}) h_{mn}(\vec{k})\ra,
= {1\over V}\int d^3\vec{r}\, e^{-i\vec{k}\cdot\vec{r}} 
\la b_i b_j b_m' b_n'\ra,
\eea
where
$h_{ij}(\vec{k})= (1/V)\int d^3\vec{x}\,  e^{-i\vec{k}\cdot\vec{x}} 
b_i(\vec{x}) b_j(\vec{x})$. 

Thus, the presence of the mean field leads to second- and 3rd-order 
statistics of the fluctuating field appearing alongside the 4th-order 
ones in the tensor $C_{ij,mn}$. Since the Stokes correlators probe the 
total field, this means that some information about the second- and 3rd-order 
statistics could be extracted from them, provided the mean field itself
can be independently determined and thus used as a ``probe'' (in fact, 
it turns out that only its orientation generally has to be known and even that 
knowledge is not always necessary, although easily obtainable; 
see \apref{ap::Obs_wmf}). 

As before, we need additional symmetry assumptions about the fluctuating 
field in order to make a transition from the Stokes correlators 
to theoretically/physically interesting quantities. 
The technically rigorous choice would be to assume that 
the statistics of $\vec{b}$ will depend on one special direction, 
that of the mean field, and be isotropic in the plane perpendicular 
to it. This, however, leads to a very large number of independent 
scalar functions appearing in the general form of 
$C_{ij,mn}$ and while it is probably worth working them all out, it is 
quite unlikely that the 6 available Stokes correlators will be sufficient 
to reconstruct anything of value. Therefore, we make a simplifying 
assumption (the physical grounds for which are discussed in \secref{sec::assumptions}) 
that the mean field is so weak ($\BB^2\ll\la b^2\ra$) that the 
fluctuating field remains statistically isotropic. 
Under this assumption, the case of a weak mean field becomes 
a straightforward generalization of the zero-mean-field case considered 
above. The main gain is that a weak mean field allows us to use 
the Stokes correlators to determine not just the power spectrum of 
the tension force but also the power spectrum of the magnetic 
field itself: as $M(k) = 4\pi k^2 c_{i,i}$ [\eqref{eq::PS}], 
it is recovered from the second-order terms in \eqref{eq::Ckgeneral}. 

\begin{figure*}
\begin{tabular}{c c c} 
{\bf MHD } & {\bf Synthetic Gaussian} \\    
\includegraphics[width=0.35\textwidth]{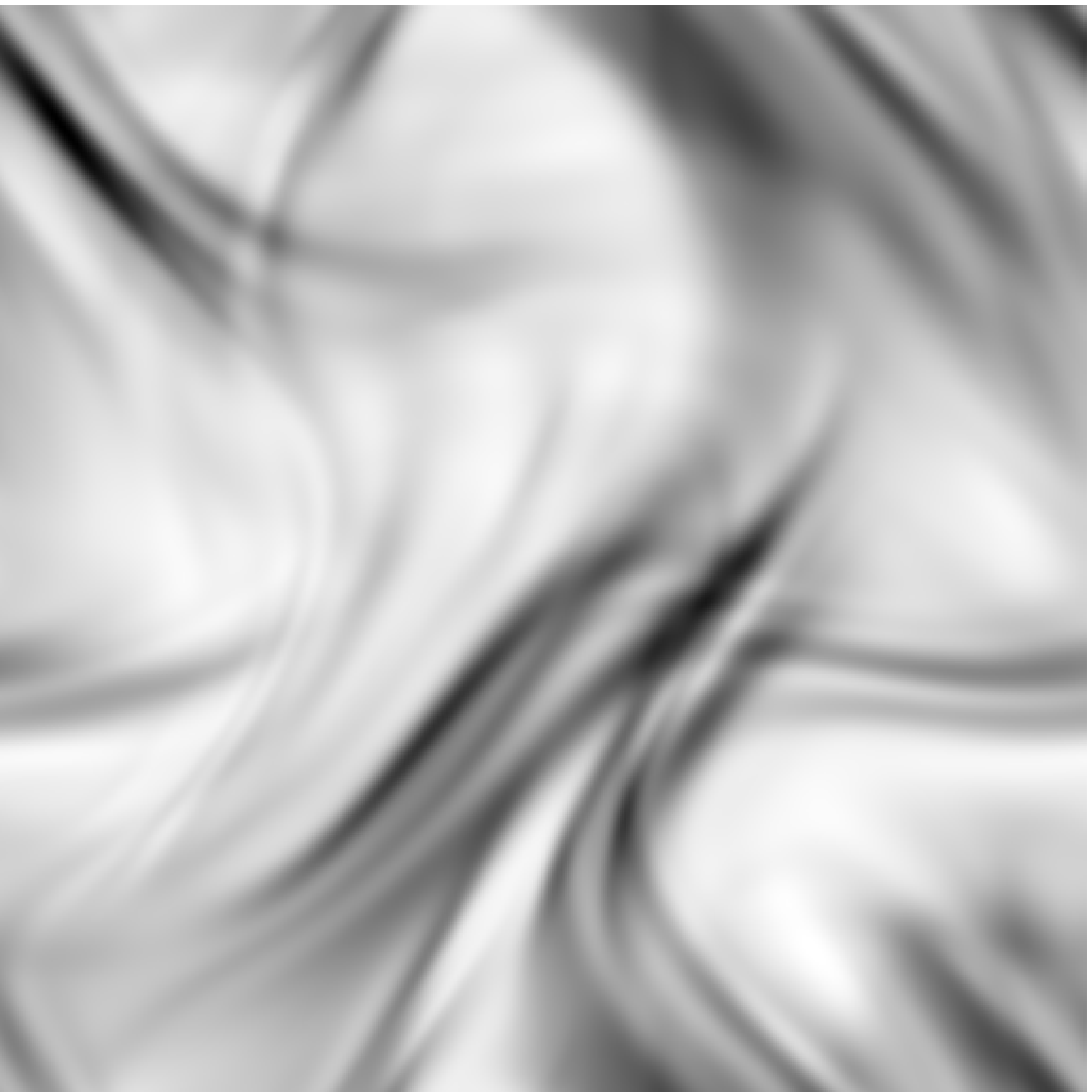} &
\includegraphics[width=0.35\textwidth]{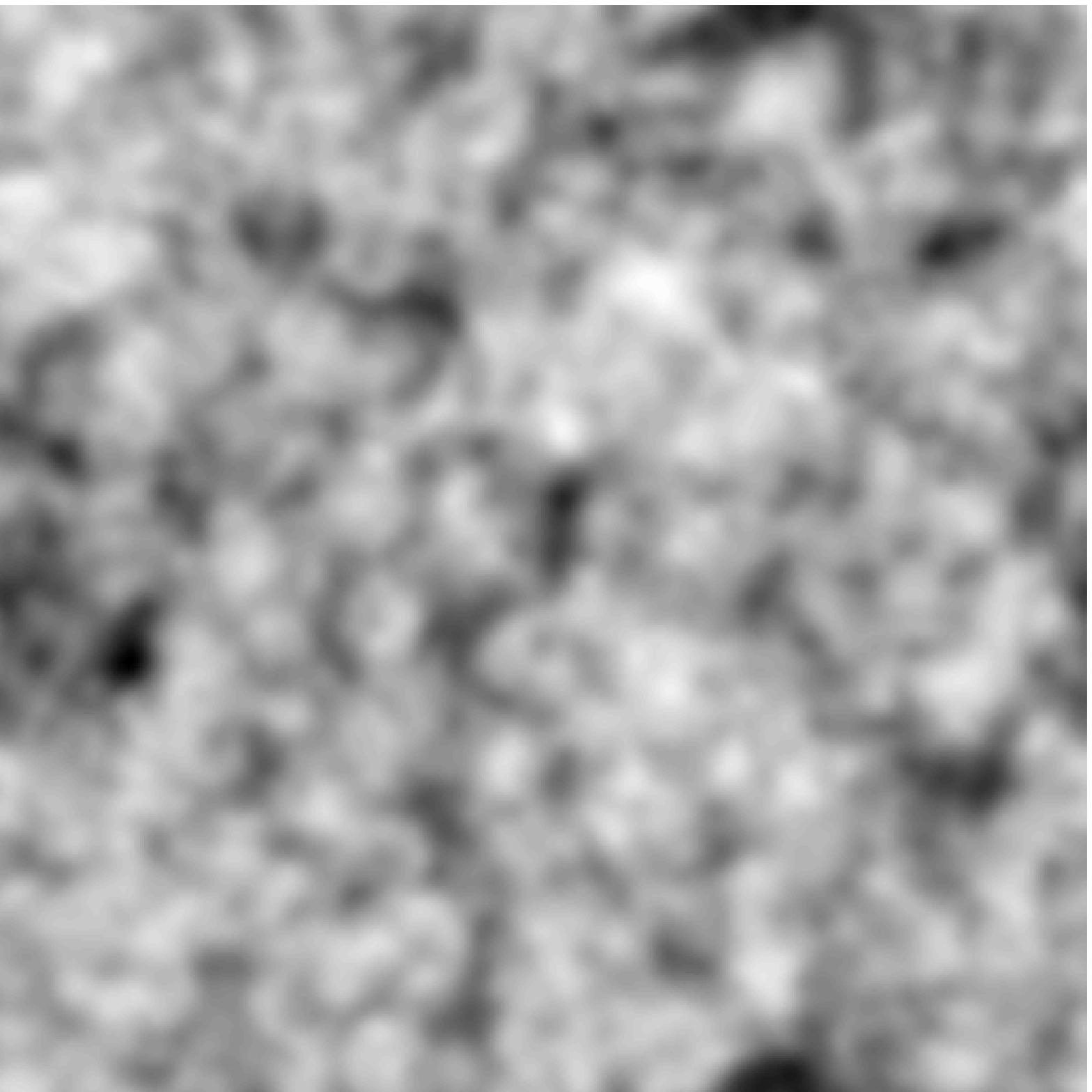} \\
$I$, $p=3$ & $I$, $p=3$\\\\
\includegraphics[width=0.35\textwidth]{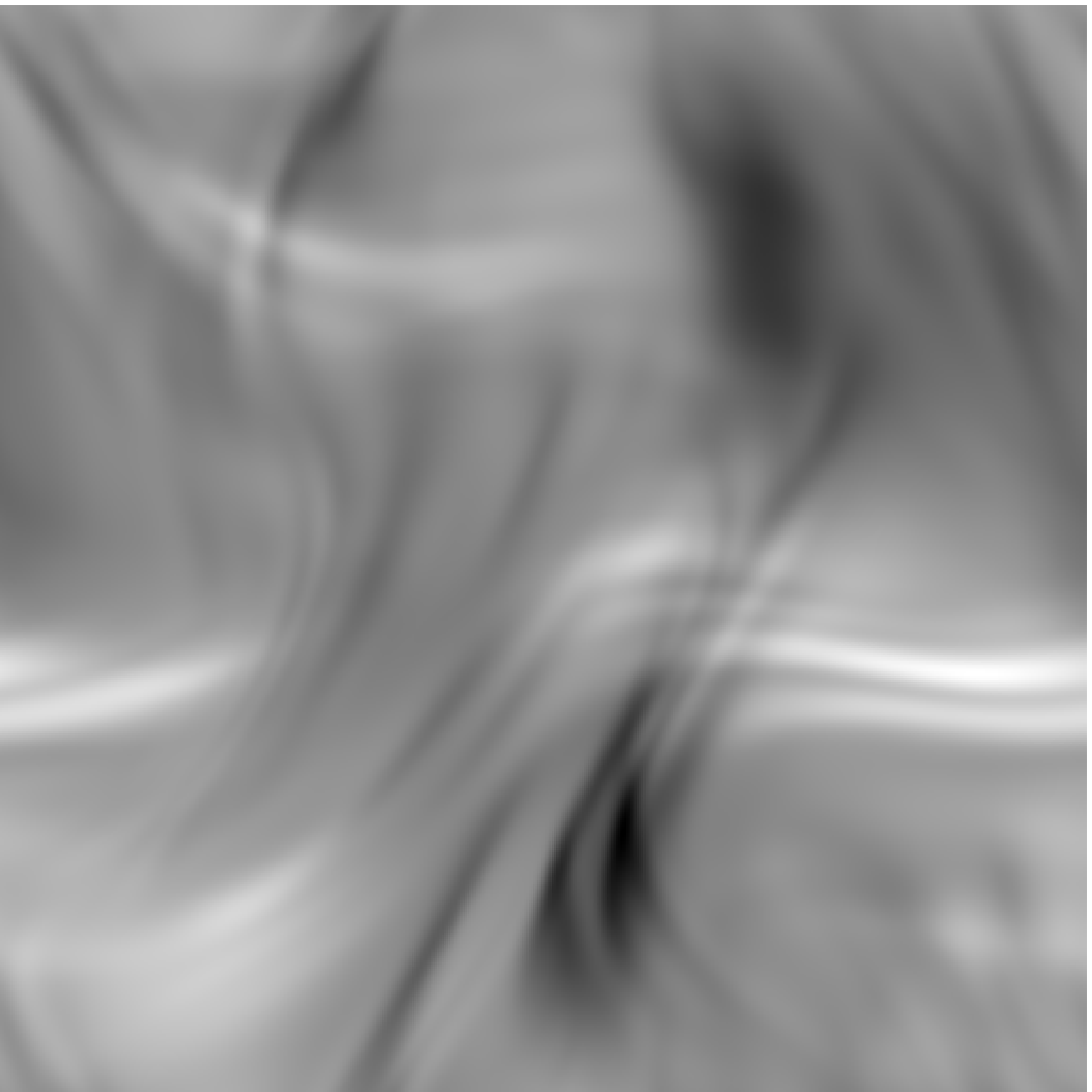} &
\includegraphics[width=0.35\textwidth]{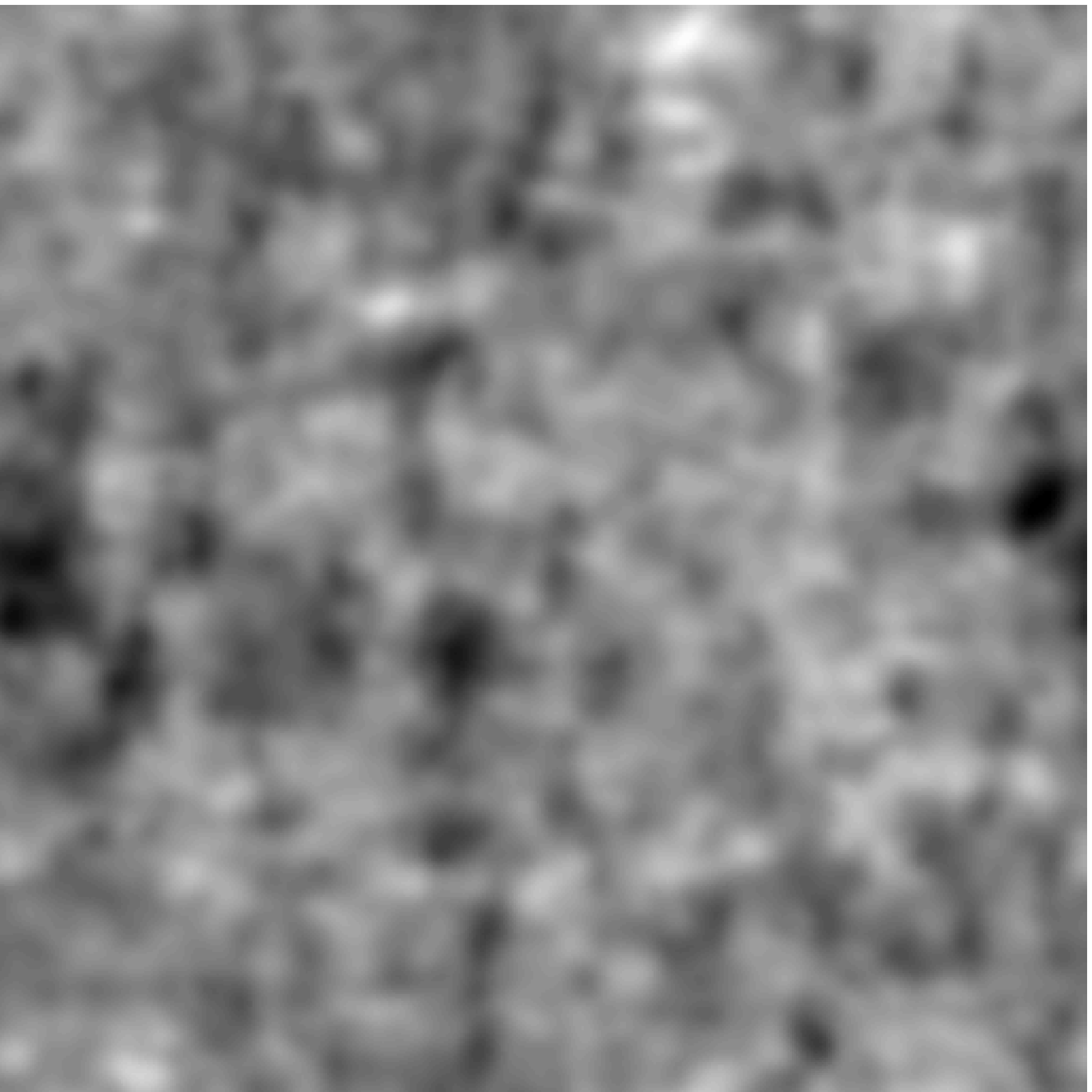} \\
$Q$, $p=3$ & $Q$, $p=3$\\\\
\includegraphics[width=0.35\textwidth]{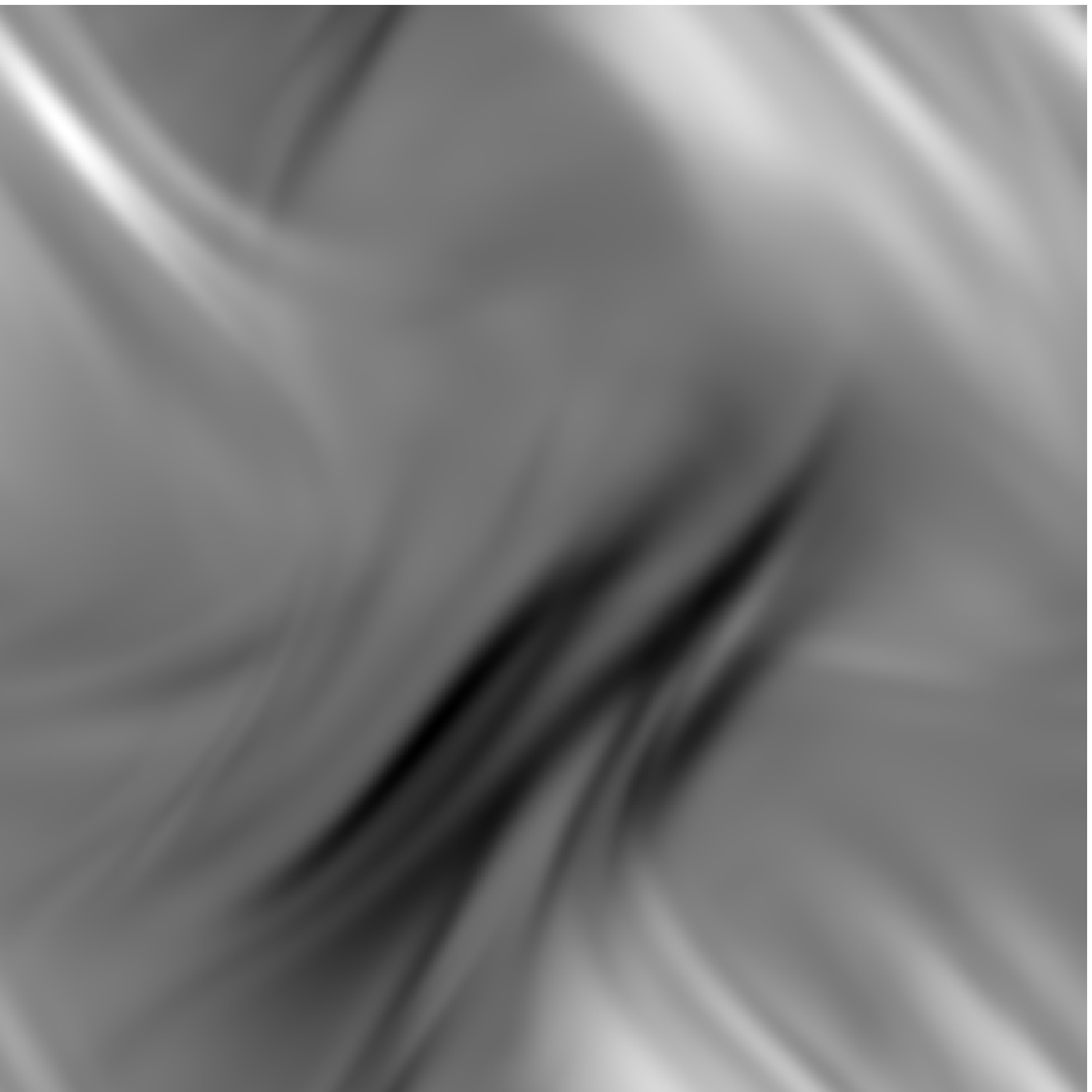} &
\includegraphics[width=0.35\textwidth]{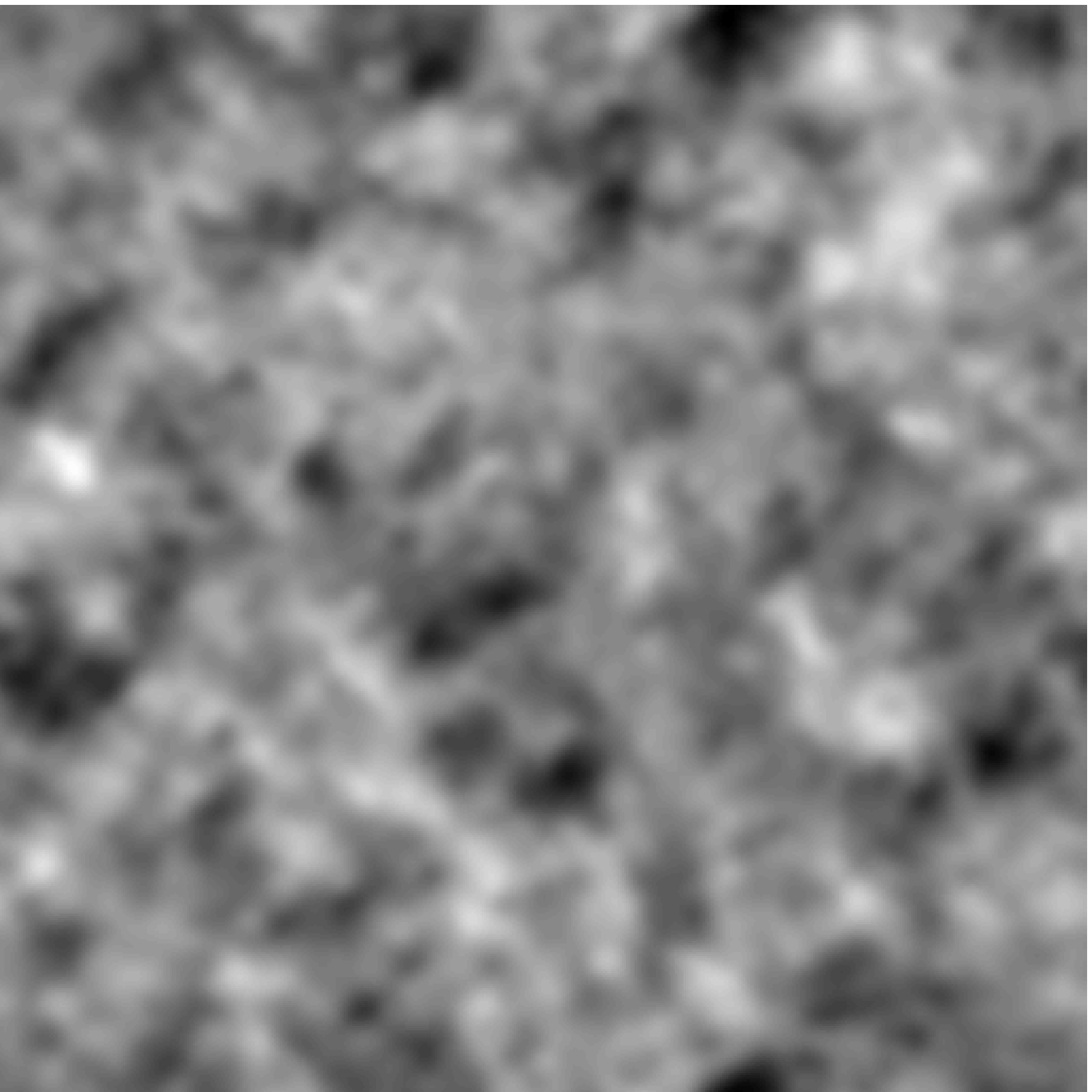}\\
$U$, $p=3$ & $U$, $p=3$
\end{tabular}
\caption{Examples of synthetic $I$, $Q$ and $U$ maps generated from 
an MHD-simulated magnetic field ({\em left panels}) 
and a synthetic Gaussian field ({\em right panels}).
The same data cubes were used as in \figref{fig::TFpoint}.}
\label{fig::maps}
\end{figure*}

\begin{figure*}
\centering
\includegraphics[angle=+90, width=0.45\textwidth]{./figs/MHDG.ps}
\includegraphics[angle=+90, width=0.45\textwidth]{./figs/Pvs3D.ps}
\caption{{\em Left panel:} The bold solid red line shows the tension-force power spectrum 
reconstructed via \eqref{eq::Tk_obs} from the synthetic Stokes maps (\figref{fig::maps}) 
based on an MHD-simulated field \citep[saturated state of the Run S4 of][]{Alex2004}. 
The bold dotted black line is the same spectrum computed directly from the full 
three-dimensional data (same as \figref{fig::spectra}, right panel). 
The errors bars on the estimated spectrum are obtained by comparing 
results from synthetic Stokes maps obtained by integrating along three 
orthogonal ``lines of sight'' (the three axes of the data cube). 
The thin solid blue line with error bars and the thin dotted black line 
represent analogous information for a synthetic Gaussian field.
{\em Right panel:} Similar to the left panel, but the reconstructed 
tension-force spectra are based not on the estimate \eqref{eq::Tk_obs} but 
on the full information about the projected (line-of-sight integrated) 
spectra, i.e., they are given by the sum of $\Phi_1$ [\eqref{eq::Phi1}] 
and $\Phi2$ [\eqref{eq::Phi2}] calculated in terms of $C_{ij,mn}(\vkp)$ 
(including its unobservable line-of-sight components).}
\label{fig::test}
\end{figure*}

The mathematical details of reconstructing the magnetic-field power 
spectrum are relegated to \apref{ap::MFSp}. 
There are many equivalent expressions that can be derived for it; 
here we display three of them: 
\bea
\nonumber
M(k) &=& {2k^2\over\BB_\perp^2\cos2\Theta} \int_0^{2\pi}d\phi\, {\rm Re}\Sigma_{IQ}\\
\nonumber
&=& {2k^2\over\BB_\perp^2\sin2\Theta} \int_0^{2\pi}d\phi\, {\rm Re}\Sigma_{IU}\\
&=& {k^2\over2\BB_\perp^2\sin2\Theta} \int_0^{2\pi}d\phi\left(\Sigma_{QQ}-\Sigma_{UU}\right),
\label{eq::Mk_obs}
\eea
where $\Theta$ is the angle between the $x$ axis and the projection of the mean field 
onto the plane perpendicular to the line of sight, $\BB_\perp$ is 
the magnitude of this perpendicular projection. Although these are 
easy to measure (\apref{ap::mf}), they are manifestly not necessary to 
determine the functional shape of the spectrum.  
Thus, we have three independent expressions from which we can deduce this functional shape. 
That the results should be consistent with one another is a good 
test of our assumptions (most importantly, the statistical isotropy of the fluctuating part 
of the field). 

The calculation of the tension-force power spectrum is entirely analogous 
to the zero-mean-field case (see \apref{ap::TenForce_wmf}). The result is 
that \eqref{eq::Tk_obs} still holds subject to two 
modifications: real part has to be taken of all Stokes correlators 
and a term proportional to $M(k)$ has to be subtracted, namely 
\bea
T_{\BB\neq0}(k) = {\rm Re} T_{\BB=0}(k) - {1\over8}\,k^2 \BB_\perp^2 M(k), 
\label{eq::Tk_obs_wmf}
\eea
where $T_{\BB=0}(k)$ is given by \eqref{eq::Tk_obs}.

Finally, a disclaimer is in order with regard to the practical applicability of 
the results obtained for the case of a weak mean field. Since we assumed the 
mean field to be small compared to the fluctuating field, $\BB^2\ll\la b^2\ra$, 
the terms in \eqref{eq::Cgeneral} that contain $\BB_i$ are small compared 
to $\la b_ib_jb'_mb'_n\ra$. Thus, in order for the second-order 
statistical information in \eqref{eq::Cgeneral} to be recoverable, 
the errors associated with the imperfect isotropy of the fluctuating 
field must be very small---smaller than $O(\BB^2)$. It is not guaranteed 
that this is either justified physically or achievable in practice and 
the verdict on the usefulness of the results of this section will depend 
on extensive numerical tests, which will not be undertaken in this paper
and are left for future work. 

\section{Numerical Tests} 
\label{sec::ASIM}

Having presented the analytical derivation of our method, we now
present a proof-of-concept numerical test by  
analyzing two data cubes containing randomly tangled magnetic fields: 
a saturated magnetic field generated by fluctuation dynamo in an MHD 
simulation (Run S4 of \citealt{Alex2004}) and a 
divergence-free, random-phased Gaussian field synthetically generated 
to have the same spectrum as the MHD field (\figref{fig::spectra}; snapshots 
of the two fields are shown in \figref{fig::TFpoint}). 
Both fields have zero mean, so the results of \secref{sec::weakmf} are 
not tested here. 

\begin{figure*}
\begin{tabular}{c c}
\includegraphics[width=0.35\textwidth]{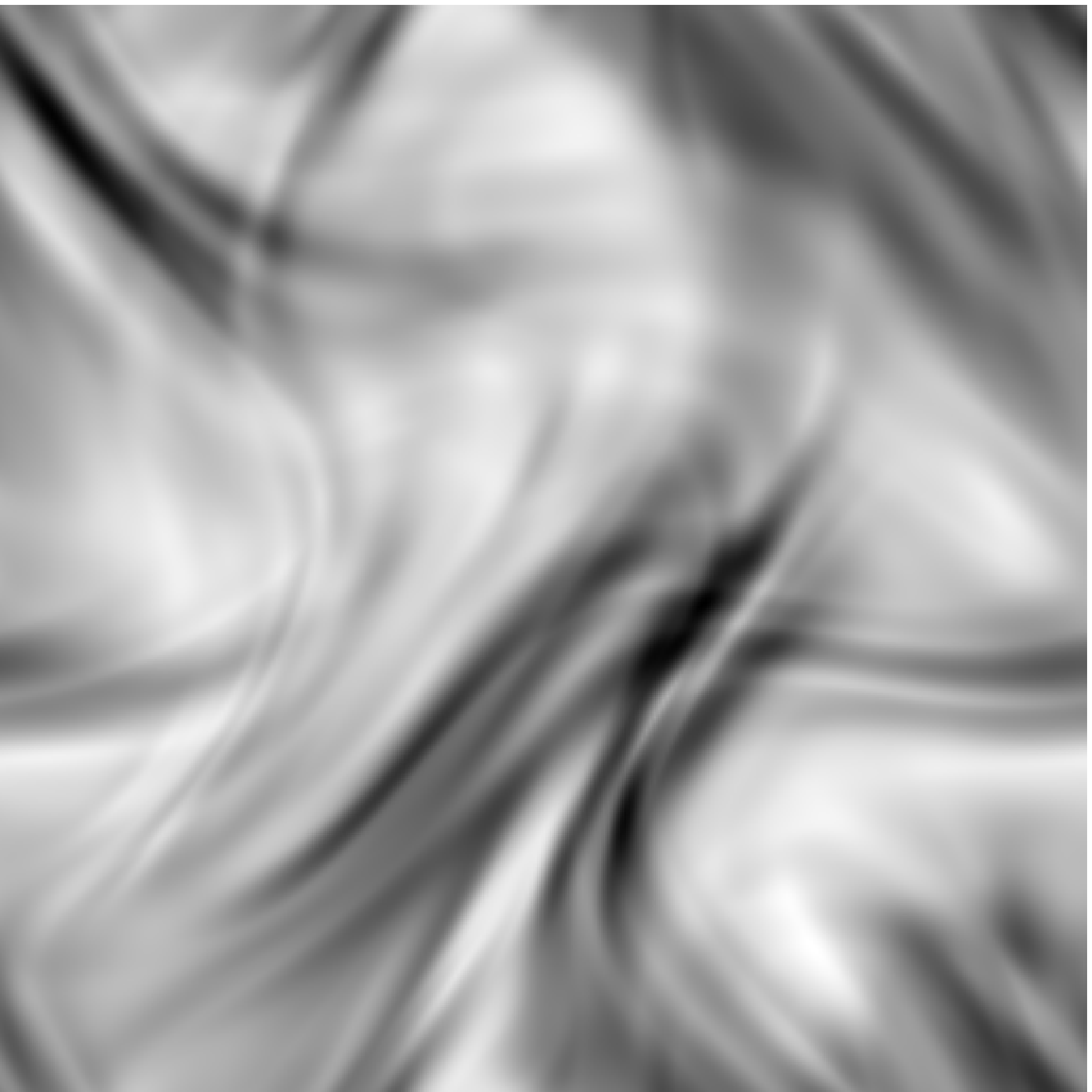} &
\includegraphics[width=0.35\textwidth]{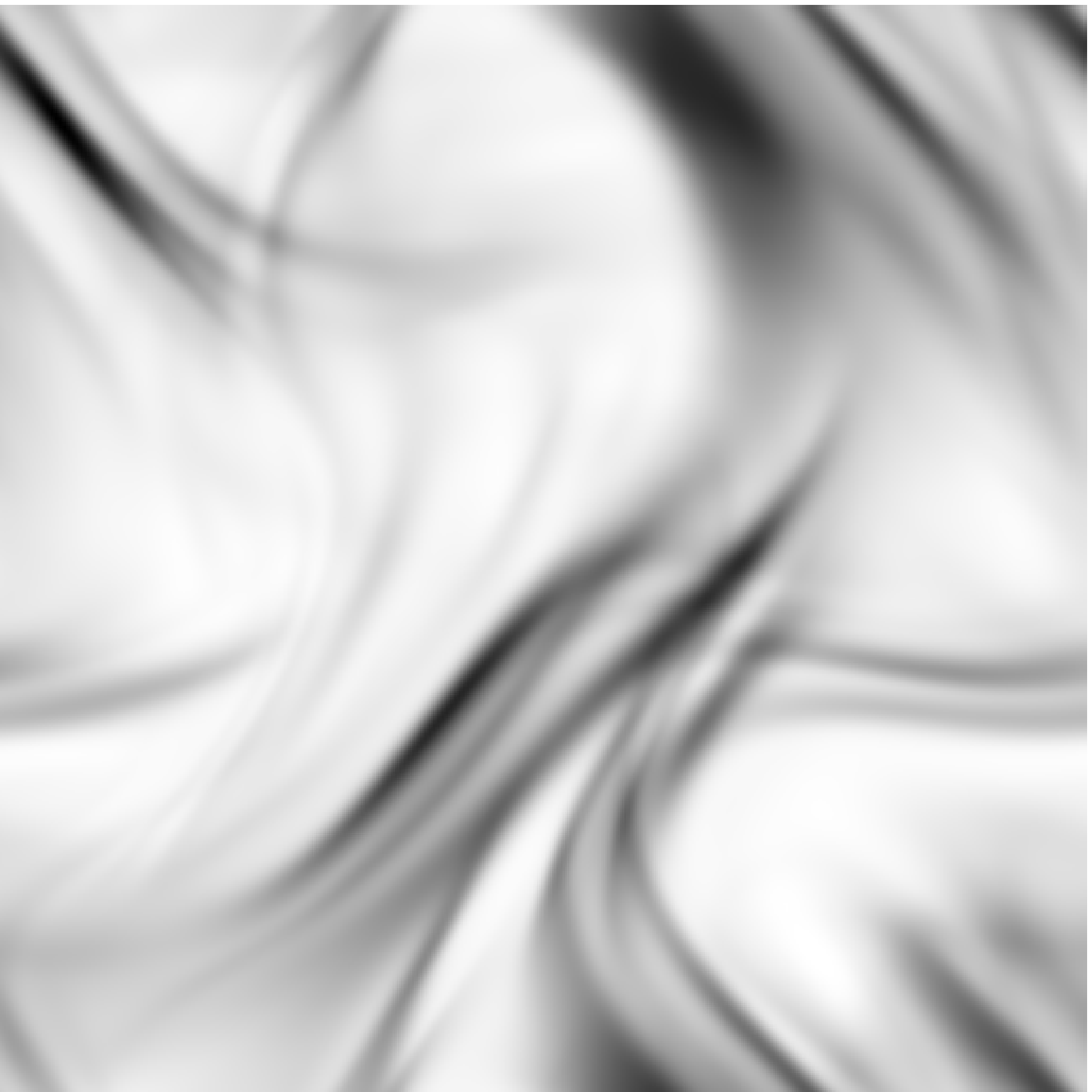}\\ 
$I$, $p=1.5$ & $I$, $p=4.5$\\\\
\includegraphics[width=0.35\textwidth]{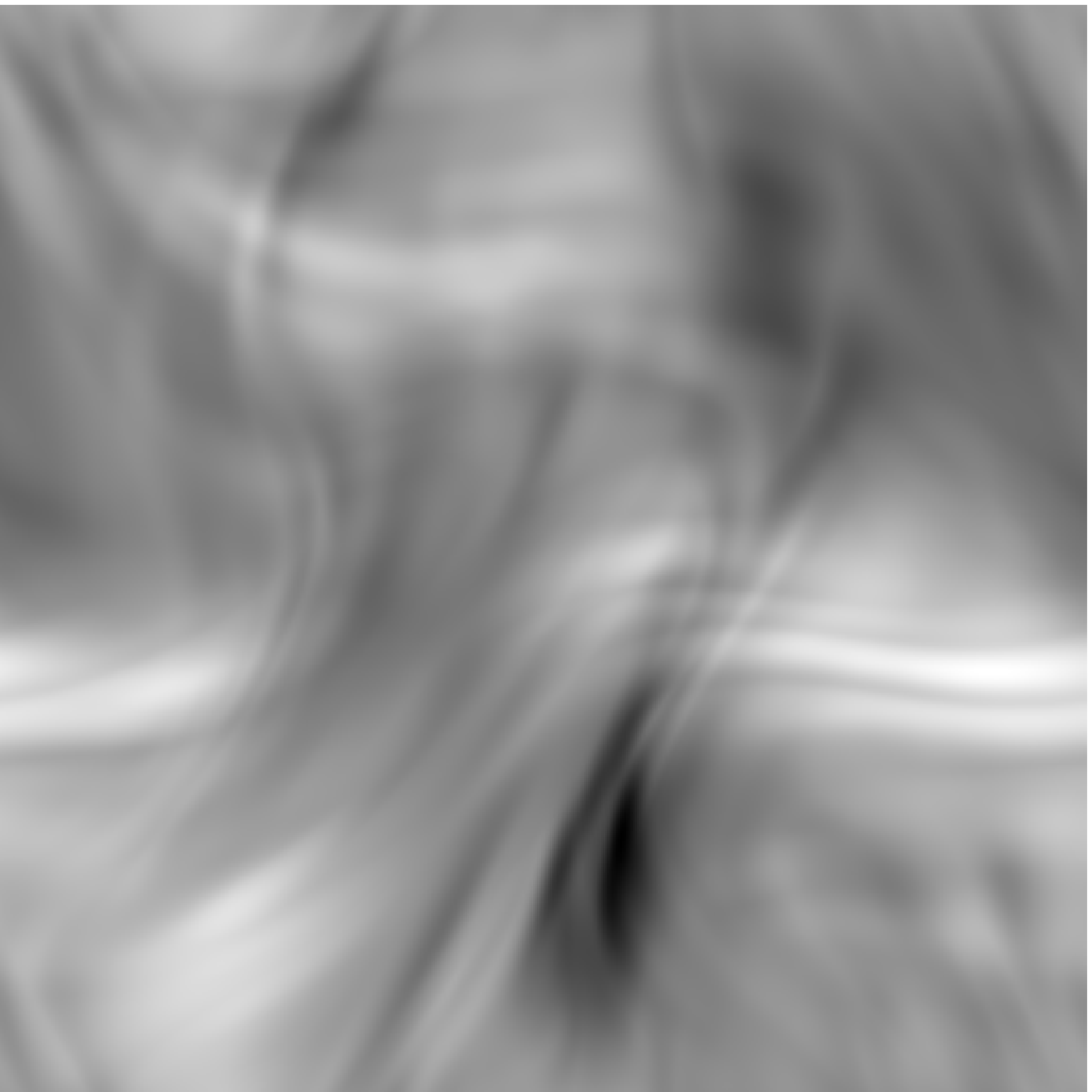} &
\includegraphics[width=0.35\textwidth]{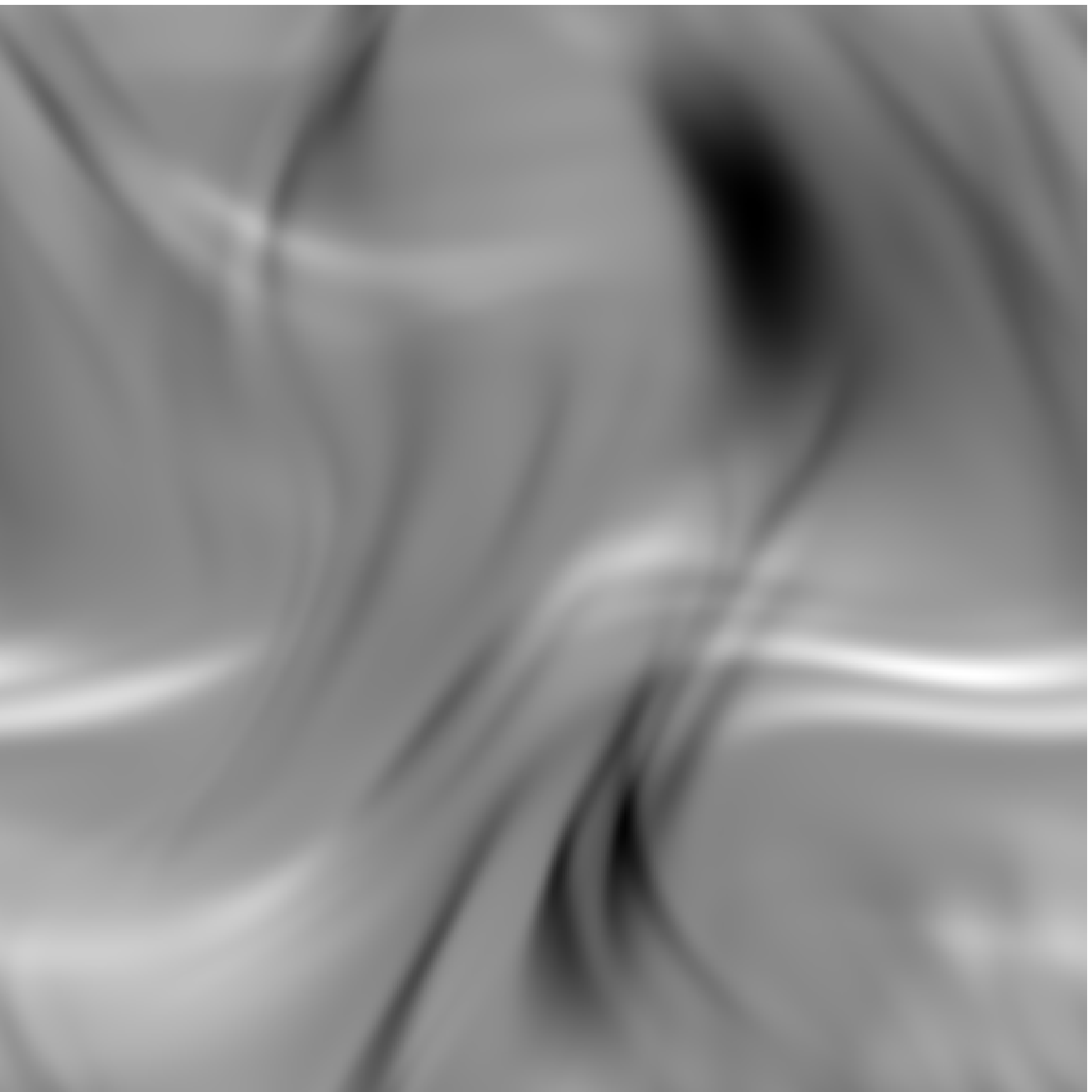}\\ 
$Q$, $p=1.5$ & $Q$, $p=4.5$\\\\
\includegraphics[width=0.35\textwidth]{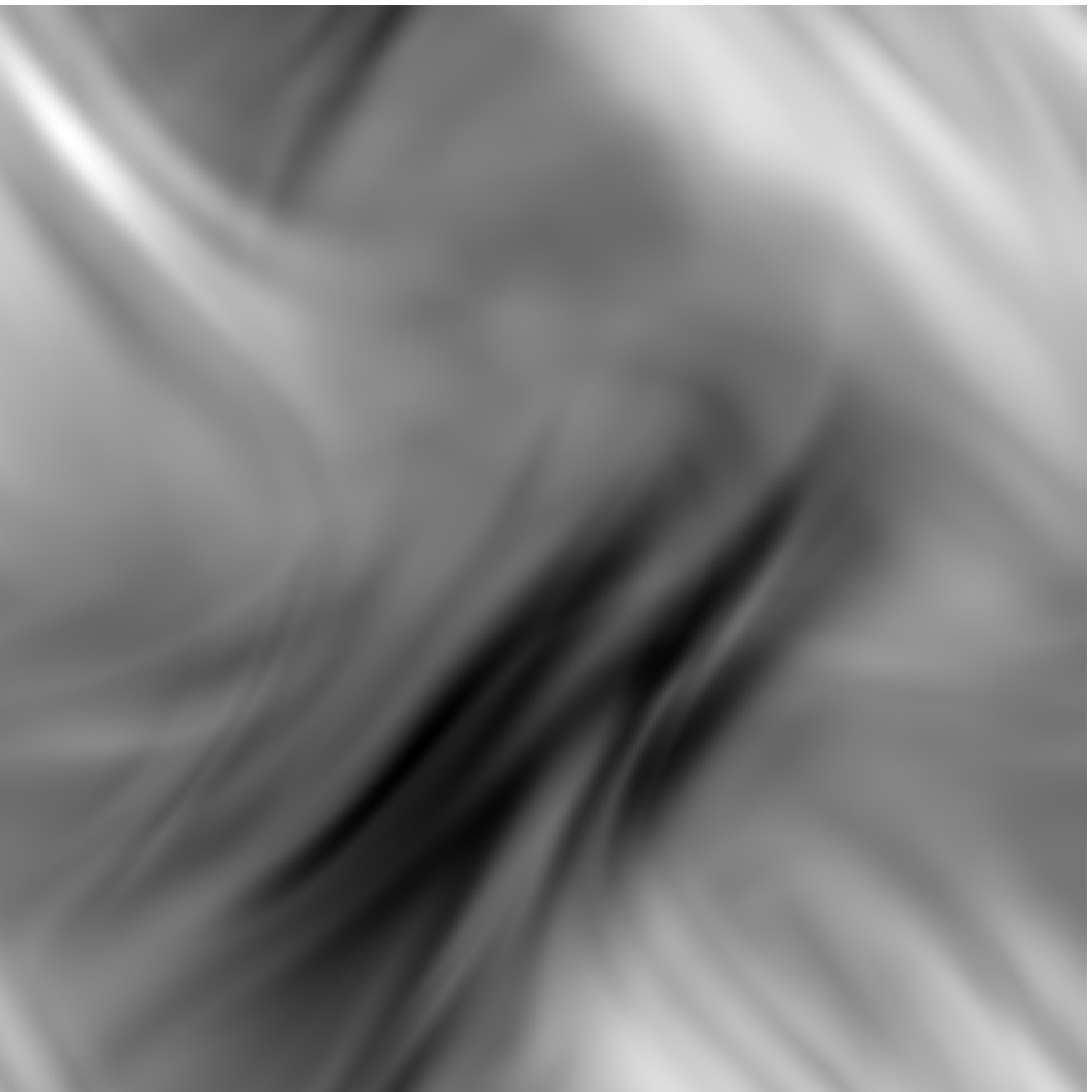} &
\includegraphics[width=0.35\textwidth]{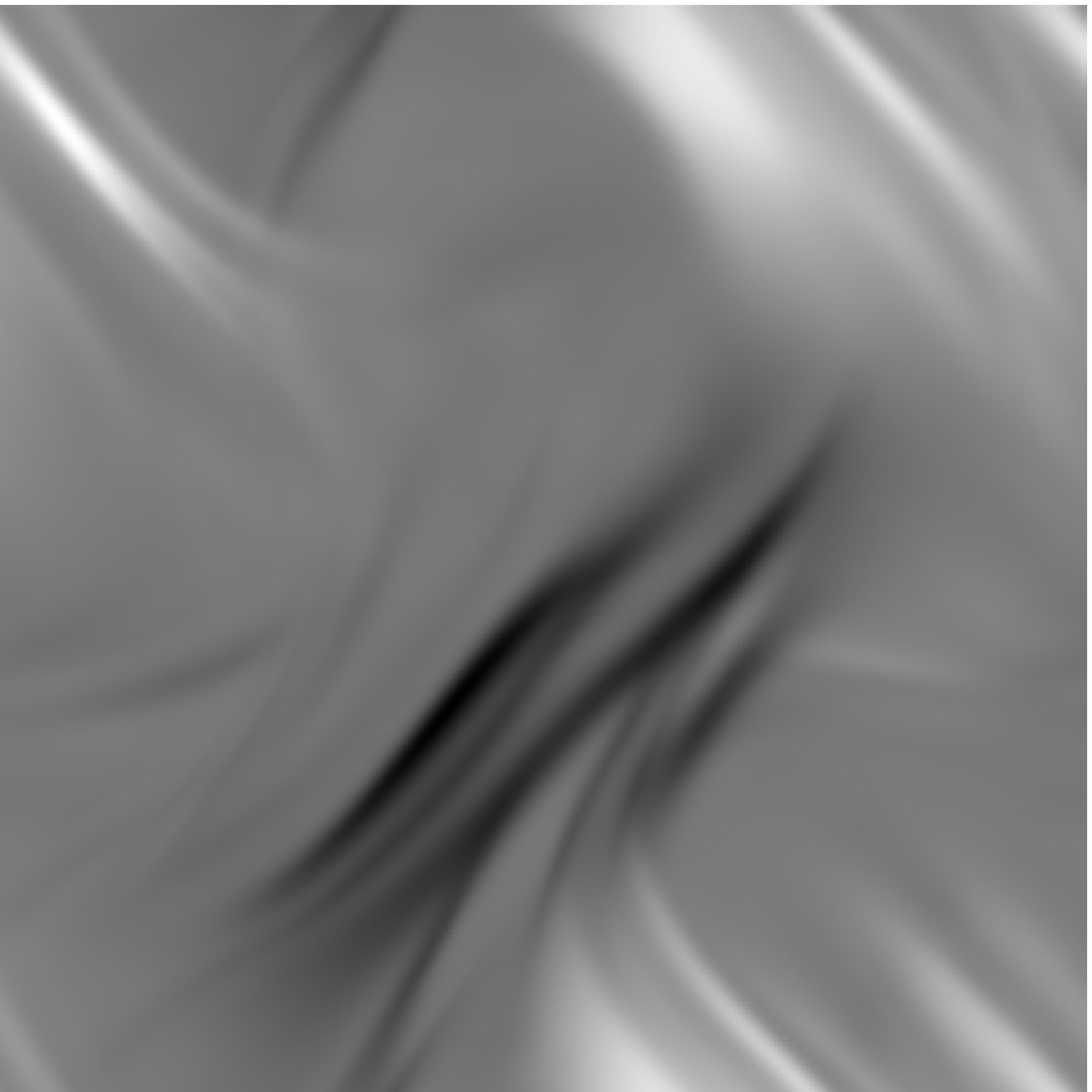}\\ 
$U$, $p=1.5$ & $U$, $p=4.5$
\end{tabular}
\caption{The Stokes maps calculated according to \eqref{eq::StoPar_p}
for two extreme values of the electron spectral index, $p=1.5$ ({\em left panels}) 
and $p=4.5$ ({\em right panels}). 
These are to be compared with the Stokes maps for $p=3$ shown in the 
left panel of \figref{fig::maps}.}
\label{fig::maps_p}
\end{figure*}

\subsection{Case of $p=3$} 

We first test the validity of our method for the case of the electron 
spectral index $p=3$, assumed throughout the analytical developments 
presented above. For each data cube, we designate one 
of its axes as the ``line of sight'' and 
integrate the field along it according to \eqref{eq::StoPar}. 
This produces three synthetic two-dimensional Stokes maps 
(\figref{fig::S}; examples of such $I$, $Q$ and $U$ maps are shown in \figref{fig::maps}). 
Since we have the full three-dimensional information for both fields, 
we can compute the tension-force power spectra directly and then compare 
them to the spectra obtained by applying our estimator, \eqref{eq::Tk_obs}, 
to the synthetic Stokes maps. 

In \figref{fig::test} (left panel) we plot the tension-force power 
spectra reconstructed using our estimator, \eqref{eq::Tk_obs}, 
for a realization of an MHD simulated field and for a synthetic 
Gaussian field. They are compared to the same spectra 
directly computed from the three-dimensional data cubes 
[according to \eqref{eq::TFPS}]. The reconstructed spectra are 
obtained as an average over three synthetic Stokes maps, each obtained 
by choosing as the ``line of sight'' one of the three orthogonal 
axes of the data cube. This allows us to estimate the accuracy of 
the reconstruction, represented in \figref{fig::test} by the error bars. 

For both types of field, the performance of our estimator is clearly 
excellent. The relative error bars for the Gaussian random field are 
substantially smaller than for the MHD field, which makes sense 
in view of the former's more small-scale and 
less structured character. 
The salient point that emerges from the comparison 
of the two fields is that the
tension-force spectrum can be recovered from the synthetic observations
with an accuracy easily permitting to discriminate between the 
structured (folded) MHD field and the structureless Gaussian one. 
This suggests that the proposed estimator is a robust tool for 
diagnosing magnetic turbulence from polarized emission data 
and for discriminating between different scenarios of magnetic-field 
evolution and saturation (see discussion in \secref{sec::why}). 

The test that we have presented only allows us to assess the quality of our
method under idealized conditions, namely, assuming that the observation
is noiseless, that no observational-window effects are present, 
that the relativistic-electron energy distribution is homogeneous and 
has the spectral index $p=3$, and that the Faraday rotation is 
either negligible or has been effectively subtracted (\secref{sec::SP}). 
Thus, the errors in our reconstructed spectra are due to two factors:  
firstly, a certain amount of information is lost in the projection 
of the three-dimensional data onto a two-dimensional Stokes map 
(the line-of-sight integration); 
secondly, the assumptions of statistical homogeneity and isotropy 
(\secref{sec::assumptions}), upon which our estimator 
depends, are imperfectly satisfied by any particular realization 
of the field. 
It is interesting to ask what is the relative contribution of these 
two sources of inaccuracy to the errors of reconstruction represented 
by the error bars in the left panel of \figref{fig::test}. 
This is addressed the right panel of the same figure, which is analogous to the 
left panel, but instead of the spectra 
reconstructed via \eqref{eq::Tk_obs}, it shows the spectra 
resulting just from the line-of-sight integration (setting $k_z=0$) 
of the full tensor $C_{ij,mn}$, i.e, they use the unobservable 
$z$ components of this tensor that enter in \eqref{eq::Phi2} 
rather than infer them from the observable components and the 
isotropy assumption. Comparing the right and left panels of 
\figref{fig::test} suggests that much of the reconstruction 
error (especially at large wave numbers) is due to the 
loss of information associated with the line-of-sight 
integration, not to imperfect isotropy---and this is despite the fact that the MHD field 
contains magnetic structures with virtually box-size parallel 
coherence lengths (see the left panel of \figref{fig::TFpoint}). 

\begin{figure*}
\begin{tabular}{c c}
\includegraphics[angle=+90, width=0.45\textwidth]{./figs/MHDG_delta_-1.5.ps} &
\includegraphics[angle=+90, width=0.45\textwidth]{./figs/MHDG_delta_-1.ps}\\
\includegraphics[angle=+90, width=0.45\textwidth]{./figs/MHDG_delta_-0.5.ps} &
\includegraphics[angle=+90, width=0.45\textwidth]{./figs/MHDG_delta_-0.25.ps}\\
\includegraphics[angle=+90, width=0.45\textwidth]{./figs/MHDG_delta_0.25.ps} &
\includegraphics[angle=+90, width=0.45\textwidth]{./figs/MHDG_delta_0.5.ps}\\
\includegraphics[angle=+90, width=0.45\textwidth]{./figs/MHDG_delta_1.ps} &
\includegraphics[angle=+90, width=0.45\textwidth]{./figs/MHDG_delta_1.5.ps}
\end{tabular}
\caption{These plots show the same comparisons as the left panel of \figref{fig::test}, 
but for a number of values of the electron spectral index $p\neq3$. 
The Stokes maps were calculated according to \eqref{eq::StoPar_p}. Some of 
these Stokes maps are shown in \figref{fig::maps_p}.}
\label{fig::test_p}
\end{figure*}

\subsection{Case of $p\neq3$}
\label{sec::spec_ind}

Assuming that the electron spectral index $p=3$ was 
an idealization of the real observational situation that we 
needed for the theoretical justification of our method because 
Stokes parameters are strictly quadratic in the magnetic field 
only if $p=3$ (see \apref{ap::SynRad}). While $p=3$ is 
not a bad approximation of reality, one cannot expect it to be 
satisfied very precisely (see discussion and references in \secref{sec::SP}),
so in order for our method to be practically useful for 
real observations, it must be reasonably insensitive to 
the exact value of $p$. This sensitivity is very easy to test. 

Let us generalize our definition of the Stokes parameters [\eqref{eq::StoPar}] 
to the case of $p\neq3$: suppressing the dimensional prefactors as before, 
we get (see \apref{ap::SynRad})
\bea
I(\vec{x}_\perp) & = & {1\over L}\int_0^L dz 
\left[ B_x^2(\vec{x}) + B_y^2(\vec{x})\right]^{(p-3)/4}
\left[ B_x^2(\vec{x}) + B_y^2(\vec{x})\right],\nonumber\\
Q(\vec{x}_\perp) & = & {1\over L}\int_0^L dz 
\left[ B_x^2(\vec{x}) + B_y^2(\vec{x})\right]^{(p-3)/4}
\left[ B_x^2(\vec{x}) - B_y^2(\vec{x})\right], \nonumber\\
U(\vec{x}_\perp) & = & {1\over L}\int_0^L dz 
\left[ B_x^2(\vec{x}) + B_y^2(\vec{x})\right]^{(p-3)/4}
2B_x(\vec{x})B_y(\vec{x}).
\label{eq::StoPar_p}
\eea 
Clearly, for $p>3$, the extra factor of $(B_x^2+B_y^2)^{(p-3)/4}$ 
causes the statistics to be effectively weighted towards 
regions where the field is stronger, for $p<3$, towards those where it is weaker. 
This point is illustrated by \figref{fig::maps_p}, which shows that 
increasing/decreasing $p$ roughly corresponds to increasing/decreasing 
the contrast in the Stokes maps. 

The range of values that $p$ can realistically be expected 
to take is roughly $p\in[1.5,3.5]$ (see references in \secref{sec::SP}). 
Since this implies that $(p-3)/4 \in [-0.375,0.125]$ are not very large powers, 
there is {\em a priori} a hope that the effect of deviations from $p=3$ might not 
be catastrophic for our estimator. This, indeed, proves to be correct. In \figref{fig::test_p}, 
we show the tension-force spectra reconstructed from Stokes maps generated using 
\eqref{eq::StoPar_p} with a number of values of $p$ 
and compare them to the true spectra. 
Even for values of $p$ significantly different 
from $3$ (roughly in the range $p\in[2.5,3.5]$), 
our estimator works extremely well, except at the 
highest wave numbers. 

Note that the extra factor of $(B_x^2+B_y^2)^{(p-3)/4}$ in \eqref{eq::StoPar_p} 
changes the overall amplitude 
of the Stokes parameters in comparison to what it would have been 
with $p=3$, so we can only hope to recover 
the functional shape of the tension-force power spectrum, not its overall 
magnitude. In the numerical data used above this potential source of reconstruction 
error is not very visible 
because values of the magnetic field are close to unity in code units, 
but in any realistic observational situation, the shift in amplitude of the 
Stokes parameters may be significant. 
Importantly, however, we see in \figref{fig::test_p} that 
in all cases we have tested, the shape of the reconstructed 
tension-force power spectrum 
still makes it unambiguously possible to discriminate between qualitatively 
different field structures as represented by the MHD and Gaussian fields.\\ 

The numerical tests presented above are meant to demonstrate 
in principle that the approach taken in this paper is a valid one. 
We did not attempt to test the robustness of our approach 
by including into our synthetic data model all of the 
complications that will arise in handling real observational data. 
A known caveat is that observational window functions, due to the finite 
size of the radio source of the telescope beam, will lead to a 
redistribution of power in the recovered spectrum, so
that the large-scale power may swamp the signal at high wave numbers
\citep{Vogt2003}. Both further tests and applications of our method to real data 
are left to future work.

\section{Conclusion}\label{sec::C}

We have demonstrated that it is possible to 
reconstruct the power spectrum of the tension force associated with 
tangled astrophysical magnetic fields as a linear combination of the radio 
synchrotron observables, the Stokes correlators. 
This was done under a set of simplifying assumptions about 
the synchrotron emission data (\secref{sec::SP}) and also by assuming 
a statistically homogeneous and isotropic stochastic magnetic field 
(\secref{sec::assumptions}). The tension-force power spectrum 
emerges as a particular case from a subset of observable 4th-order 
statistics (\secref{sec::TenForce} and \apref{ap::TenForce})---a nontrivial 
fact because in general, the Stokes maps do not carry sufficient 
information to reconstruct all of the 4th-order correlators of 
the magnetic field (see \secref{sec::SC} and \apref{ap::GenObs}). 

The observability of the tension-force power spectrum is 
a stroke of good fortune because this quantity plays an important role 
in diagnosing the spatial structure of the magnetic turbulence 
\citep{Alex2004} and allows one to distinguish between different 
theoretical scenarios for the evolution and saturation of the cosmic 
magnetic field, which was not possible to do on the basis of  
lower-order statistics such as the magnetic power spectrum;
it also reveals physically interpretable dynamical properties of the 
system under observation, namely the force exerted by the field 
on the ambient plasma (see discussion in \secref{sec::why}).

Furthermore, we have shown that if the observed magnetic field 
possesses a small regular component that does not affect the isotropy 
of the fluctuating part of the field, it may be 
possible to obtain from the Stokes maps the power spectrum of 
the fluctuating field itself, as well as that of its tension force
(\secref{sec::weakmf}). 

Thus, physically relevant information about the spatial structure and 
dynamical properties
of the magnetic turbulence is contained in the polarized emission 
maps and can be extracted. This work is an attempt to pave
the way towards analyzing the large amount of existing and
upcoming radio-synchrotron observational data \citep[see, e.g.,][]{Gaensler2006,Ensslin2006,Beck2008} 
with the aim of achieving a better understanding of the nature of magnetized 
turbulence in cosmic plasmas. 

\section*{Acknowledgements}
AHW would like to thank Tarek Yousef and Martin Reinecke. 
He acknowledges travel support from the Leverhulme 
Trust International Network for Magnetized Plasma Turbulence.
The work of AAS was supported in part by a PPARC/STFC Advanced Fellowship 
and by the STFC Grant ST/F002505/1. 

\bibliography{SC}
\bibliographystyle{mn2e}


\appendix 

\section[]{Synchrotron Emission and the Stokes Parameters}
\label{ap::SynRad}

A spatially homogeneous, pitch-angle-isotropic and power-law distributed 
in energy relativistic-electron population is assumed [\eqref{eq::eldistr}], 
The resulting synchrotron emission is partially linearly polarized. 
Its intensity and polarization depend 
solely on the magnitude and orientation of the magnetic field 
$\vec{B}_\perp$ projected onto the plane perpendicular to the line of sight 
and on the electron distribution [\eqref{eq::eldistr}]. 

The synchrotron emissivity (i.e., power per unit volume per frequency per solid
angle) is usually subdivided into two components, respectively
perpendicular and parallel to $\vec{B}_\perp$: following \citet{RL},
\bea
\nonumber
j_\perp (\omega,\vec{x}) &=& \left[F(p) + G(p)\right]\omega^{(1-p)/2} 
|\vec{B}_\perp(\vec{x})|^{(p+1)/2}\\
j_\parallel (\omega,\vec{x}) &=& \left[F(p) - G(p)\right]\omega^{(1-p)/2} 
|\vec{B}_\perp(\vec{x})|^{(p+1)/2},
\eea
where $\omega = 2\pi\nu$, $\nu$ is the observation frequency, $\vec{x}$ is the spatial position,
$p$ is the spectral index of the electron distribution [\eqref{eq::eldistr}] and 
\bea
\nonumber
F(p) &=& \frac{\sqrt{3}\,e^3}{32\pi^2 m_ec^2} 
\left(\frac{2m_ec}{3e}\right)^{(1-p)/2} C\\
\nonumber 
&&\times\,\, \Gamma\left(\frac{p}{4}-\frac{1}{12}\right) 
\frac{2^{(p+1)/2}}{p+1} \Gamma\left(\frac{p}{4}+\frac{19}{12}\right),\\ 
\nonumber
G(p) &=& \frac{\sqrt{3}\,e^3}{32\pi^2 m_ec^2} 
\left(\frac{2m_ec}{3e}\right)^{(1-p)/2} C\\
&&\times\,\, \Gamma\left(\frac{p}{4}-\frac{1}{12}\right) 
2^{(p-3)/2} \Gamma\left(\frac{p}{4}+\frac{7}{12}\right),
\eea
where $m_e$ is the electron mass, $e$ is its charge, $c$ is the speed of light, 
and $C$ is the prefactor of the electron distribution [\eqref{eq::eldistr}].

The specific intensity ${\rm I}$ and the polarized specific intensity ${\rm PI}$ 
are given by the following line-of-sight integrals \citep[see][]{Burn1966}:
\bea
\nonumber
{\rm I}(\omega,\vec{x}_\perp) &=& \int_{\rm there}^{\rm here} dz 
\left[j_\perp(\omega,\vec{x}) + j_\parallel(\omega,\vec{x})\right],\\
{\rm PI}(\omega,\vec{x}_\perp) &=& \int_{\rm there}^{\rm here} dz 
\left[j_\perp(\omega,\vec{x}) - j_\parallel(\omega,\vec{x})\right]
e^{-2i\chi(\vec{x})},
\eea
where $z$ is the line-of-sight coordinate, 
$\vec{x}_\perp = (x,y)$ is the position vector in the plane of the sky 
(perpendicular to the line of sight) and the polarization angle is given by 
\bea
\chi(\vec{x}) = \chi_0(\vec{x}) + \lambda^2{\rm RM}(\vec{x}), 
\label{eq::chi}
\eea
where $\lambda=c/\nu$ is wavelength of the observed emission, 
the intrinsic polarization angle is
\bea
\chi_0(\vec{x}) = \tan^{-1} {B_y\over B_x},
\eea
and the Faraday rotation measure
\bea
{\rm RM}(\vec{x}) = {e^3\over 2\pi m_e^2 c^4} 
\int_{\rm there}^{\rm here} dz\, n_{{\rm th}e}(\vec{x}) B_z(\vec{x}),
\eea
where $n_{{\rm th}e}$ is the density of thermal electrons and $B_z$ the 
projection of the magnetic field on the line of sight.
 
The Stokes parameters are now defined as follows
\bea
I = \int d\Omega \, {\rm I},\qquad
Q-iU = \int d\Omega \, {\rm PI},
\eea
where the integration is over the solid angle of the angular
resolution element (the observational beam). 
If the spectral index is taken to be $p=3$
and the Faraday rotation in \eqref{eq::chi} is assumed to be negligible
(as discussed in \secref{sec::SP}), the Stokes parameters depend 
quadratically on the components of the magnetic field 
perpendicular to the line of sight, $B_x = B_\perp\cos\chi_0$ 
and $B_y = B_\perp\sin\chi_0$. Indeed, using the above definitions, we get
\bea
\nonumber
I &=& 2F(3)\, \omega^{-1}
\int d\Omega \int_{\rm there}^{\rm here} dz \left(B_x^2 + B_y^2\right),\\
\nonumber
Q &=& 2G(3)\, \omega^{-1}
\int d\Omega \int_{\rm there}^{\rm here} dz \left(B_x^2 - B_y^2\right),\\
U &=& 2G(3)\, \omega^{-1}
\int d\Omega \int_{\rm there}^{\rm here} dz\, 2B_x B_y.
\eea
From these formulae, we recover the analytically convenient 
definitions of the Stokes parameters, \eqref{eq::StoPar}, 
by dropping the dimensional prefactors and the integration 
over the angular resolution element and normalizing 
the integrals by the depth of the emission region.  

Somewhat more generally, for arbitrary $p$, but still neglecting 
the Faraday rotation, we have 
\bea
\nonumber
I &=& 2F(p)\, \omega^{(1-p)/2}
\int d\Omega \int_{\rm there}^{\rm here} dz \left(B_x^2 + B_y^2\right)^{(p-3)/4}
\left(B_x^2 + B_y^2\right),\\
\nonumber
Q &=& 2G(p)\, \omega^{(1-p)/2}
\int d\Omega \int_{\rm there}^{\rm here} dz \left(B_x^2 + B_y^2\right)^{(p-3)/4}
\left(B_x^2 - B_y^2\right),\\
\nonumber
U &=& 2G(p)\, \omega^{(1-p)/2}
\int d\Omega \int_{\rm there}^{\rm here} dz \left(B_x^2 + B_y^2\right)^{(p-3)/4}
2B_x B_y.\\
\eea
These formulae are the basis for \eqref{eq::StoPar_p}. 

\section{Fourth-Order Correlation Tensor 
and Its Representation in Terms of Stokes Correlators}
\label{ap::theory}

In this Appendix, we derive the general form of the 4th-order correlation 
tensor $C_{ij,mn}$ [\eqref{eq::Cijmn}] for a statistically homogeneous and isotropic 
magnetic field and show what part of the relevant statistical information 
can be recovered using Stokes correlators. 

\subsection{Symmetries and the General Form of $C_{ij,mn}$}

In \eqref{eq::Ckgeneral}, the tensor $C_{ij,mn}$ is written in Fourier space 
in terms of the mean field $\BB_i$ and of the second-, 3rd-, and 4th-order 
correlation tensors of the fluctuating field $b_i$, denoted 
$c_{i,m}$, $c_{ij,m}$ and $c_{ij,mn}$.  
Each of these correlation tensors depends on a certain number of scalar 
correlation functions \citep[see, e.g.,][]{Robertson1940}. 
This number can be constrained if we take into account 
some intrinsic properties of correlation tensors (permutation of indices), 
of the field they are constructed from (it is a real, divergence-free field), 
and additional symmetries we assume (homogeneity and isotropy). 
Let us implement these constraints.
The procedure is least cumbersome when applied to the second-order 
correlation tensor. We will explain it in detail on this example and then 
proceed analogously with the 3rd- and 4th-order correlators. 
All further calculations will be in Fourier space, but 
exactly analogous calculations can be done in position space if 
it is necessary to compute position-space correlators. 

\subsubsection{Second-Order Correlation Tensor}

For a statistically isotropic field, the second-order correlation 
tensor depends on three scalar functions---this is shown by constructing 
$c_{i,m}$ out of all possible isotropic second-rank tensors. In three dimensions, 
the available building blocks for these tensors are $\delta_{im}$, 
$\eps_{imp}$ and $\kk_i$, the unit vector in the direction of $\vec{k}$. 
Therefore, 
\bea
c_{i,m}(\vec{k}) = {1\over2}\left[m_1\delta_{im} + m_2\kk_i\kk_m\right] + i m_3 \eps_{imp}\kk_p,
\label{eq::c2gen}
\eea 
where the scalar coefficients $m_1$, $m_2$, $m_3$ can only depend on $k=|\vec{\vec{k}}|$. 

Since $b_i(\vec{k})$ is 
a Fourier transform of a real function, we must have $b_i(-\vec{k})=b_i^*(\vec{k})$, 
whence
\bea
c_{i,m}(-\vec{k}) = c_{i,m}^*(\vec{k}). 
\eea
It is easy to see that this implies that $m_1$, $m_2$ and $m_3$ are real (the factor 
of $i$ in front of $m_3$ was chosen deliberately to arrange for this outcome). 

Since $c_{i,m}$ is a correlation tensor, it has a symmetry with respect to 
permutation of its indices: 
\bea
c_{m,i}(\vec{k}) = \la b_m^*(\vec{k}) b_i(\vec{k})\ra = c_{i,m}^*(\vec{k}). 
\eea
This does not bring any new information beyond the reality of $m_1$, $m_2$ and $m_3$. 

Finally, the magnetic field is solenoidal, $k_i b_i(\vec{k})=0$, so we must have
\bea
k_i c_{i,m} = k_m c_{i,m} = 0.
\label{eq::sol2}
\eea
This gives $m_2=-m_1$, so the general form of the second-order correlation tensor is 
\bea
c_{i,m}(\vec{k}) = {1\over2}\,m_1(k) \left(\delta_{im} - \kk_i\kk_m\right) + i m_3(k)\eps_{imp}\kk_p, 
\label{eq::c2_gen}
\eea
i.e., it depends only on two scalar functions. If we take the trace of this tensor, 
we obtain the magnetic-energy power spectrum [\eqref{eq::PS}]:
\bea
M(k) = 4\pi k^2 c_{i,i}(k) = 4\pi k^2 m_1(k),
\label{eq::Mdef}
\eea
so we do not need to know $m_3$ if we are only interested in the power spectrum. 
\citet{Vogt2003,Vogt2005} used this property to propose a way to measure the magnetic power 
spectrum solely in terms of the scalar correlation function of the Faraday rotation 
measure associated with a given magnetic-field distribution: although only one scalar 
function was available this way, assuming isotropy and restricting one's attention to a particular 
quantity of physical interest made it possible to make do with incomplete information. 
We follow the same basic philosophy in this paper, primarily as applied to the 
4th-order statistics. 

Note that $m_3$ is a measure of reflection (parity, or mirror) non-invariance of 
the magnetic field. If $m_3\neq0$, the field has helicity.   
If we demand mirror symmetry of the field, 
\bea
c_{i,m}(-\vec{k}) = c_{i,m}(\vec{k}), 
\eea
we find $m_3=0$. We will see that normally we do not have to make this 
assumption because in many cases, the mirror-noninvariant 
terms are not present in the quantities of interest (as was the case 
with the power spectrum). 

\subsubsection{Third-Order Correlation Tensor}

Analogously to the above, we construct the general isotropic 3rd-order tensor 
as follows
\bea
\nonumber
c_{ij,m}(\vec{k}) &=& i\left(a_1 \delta_{ij}\kk_m + a_2 \delta_{im}\kk_j + a_3 \delta_{jm}\kk_i 
+ a_4\kk_i\kk_j\kk_m\right) + a_5\eps_{ijm} \\
&&\,\,+ a_6\eps_{ijp}\kk_p\kk_m + a_7\eps_{imp}\kk_p\kk_j + a_8\eps_{jmp}\kk_p\kk_i,  
\eea
where $a_1$, \dots, $a_8$ are functions of $k=|\vec{k}|$ only. 

Reality of the fields $h_{ij}$ and $b_m$ implies
\bea
c_{ij,m}(-\vec{k}) = c_{ij,m}^*(\vec{k}),
\eea
whence $a_1$, \dots, $a_8$ are all real. 

Permutation symmetry, 
\bea
c_{ji,m}(\vec{k}) = c_{ij,m}(\vec{k}),
\eea
implies $a_2=a_3$, $a_5=a_6=0$, and $a_7=a_8$. 

Solenoidality of the magnetic field implies 
\bea
k_m c_{ij,m}(\vec{k}) = 0,
\label{eq::sol3}
\eea
whence $a_1=0$ and $a_4=-2a_2$. 

Thus, the general form of the 3rd-order correlation tensor is 
\bea
\nonumber
c_{ij,m}(\vec{k}) &=& ia_2(k)\left(\delta_{im}\kk_j + \delta_{jm}\kk_i - 2\kk_i\kk_j\kk_m\right)\\ 
&&+\,\, a_7(k)\left(\eps_{imp}\kk_p\kk_j + \eps_{jmp}\kk_p\kk_i\right).
\label{eq::c3gen}
\eea

\subsubsection{Fourth-Order Correlation Tensor}
\label{ap::4thorder}

In the 4th order, the number of terms in the general tensor becomes quite large. 
Constructing this general form out of the usual building blocks, 
$\delta_{ij}$, $\kk_i$ and $\eps_{ijm}$, we get, 
via straightforward combinatorics, 
\bea
\nonumber
c_{ij,mn}(\vec{k}) &=& 
c_1\delta_{ij}\delta_{mn} + c_2\delta_{im}\delta_{jn} + c_3\delta_{in}\delta_{jm}\\
&&+\,\, c_4\delta_{ij}\kk_m\kk_n + c_5\delta_{mn}\kk_i\kk_j
\nonumber\\ 
&&+\,\, c_6\delta_{im}\kk_j\kk_n + c_7\delta_{in}\kk_j\kk_m 
+ c_8\delta_{jm}\kk_i\kk_n + c_9\delta_{jn}\kk_i\kk_m
\nonumber\\
&&+\,\, c_{10}\kk_i\kk_j\kk_m\kk_n
\nonumber\\
&&+\,\, i\left( c_{11}\eps_{ijm}\kk_n + c_{12}\eps_{ijn}\kk_m 
+ c_{13}\eps_{imn}\kk_j + c_{14}\eps_{jmn}\kk_i\right.
\nonumber\\
&&+\,\, c_{15}\eps_{ijp}\kk_p\delta_{mn} + c_{16}\eps_{mnp}\kk_p\delta_{ij} 
+c_{17}\eps_{imp}\kk_p\delta_{jn} 
\nonumber\\
&&+\,\, c_{18}\eps_{inp}\kk_p\delta_{jm}
+ c_{19}\eps_{jmp}\kk_p\delta_{in} + c_{20}\eps_{jnp}\kk_p\delta_{im}
\nonumber\\
&&+\,\, c_{21}\eps_{ijp}\kk_p\kk_m\kk_n + c_{22}\eps_{mnp}\kk_p\kk_i\kk_j
+ c_{23}\eps_{imp}\kk_p\kk_j\kk_n
\nonumber\\
&&+\left. c_{24}\eps_{inp}\kk_p\kk_j\kk_m
+ c_{25}\eps_{jmp}\kk_p\kk_i\kk_n + c_{26}\eps_{jnp}\kk_p\kk_i\kk_m\right),
\nonumber\\
\eea
where $c_1$, \dots, $c_{26}$ are functions of $k=|\vec{k}|$ only. 
Note that there are no terms of the form $\eps_{ijp}\kk_p\eps_{mnq}\kk_q$ because 
\bea
\nonumber
\eps_{ijp}\eps_{mnq} &=& \delta_{im}\delta_{jn}\delta{pq} 
+ \delta_{in}\delta{jq}\delta_{pm}
+ \delta_{iq}\delta{jm}\delta_{pn}\\
&&-\,\, \delta_{im}\delta_{jq}\delta_{pn}
-\delta_{in}\delta_{jm}\delta{pq}
-\delta_{iq}\delta_{jn}\delta_{pm}, 
\eea
so such terms are already present in the general form we have constructed. 

Reality of the field $h_{ij}=b_ib_j$ implies
\bea
c_{ij,mn}(-\vec{k}) = c_{ij,mn}^*(\vec{k}),
\eea
whence $c_1$, \dots, $c_{26}$ are all real. 

There are three permutation symmetries:
\bea
c_{ji,mn}(\vec{k}) = c_{ij,mn}(\vec{k})
\eea
gives $c_2=c_3$, $c_6=c_8$, $c_7=c_9$, $c_{11}=c_{12}=c_{15}=c_{21}=0$, 
$c_{13}=c_{14}$, $c_{17}=c_{19}$, $c_{18}=c_{20}$, $c_{23}=c_{25}$, $c_{24}=c_{26}$, 
\bea
c_{ij,nm}(\vec{k}) = c_{ij,mn}(\vec{k}),
\eea
gives additionally $c_6=c_7$, $c_8=c_9$, $c_{13}=c_{14}=c_{16}=c_{22}=0$, 
$c_{17}=c_{18}$, $c_{19}=c_{20}$, $c_{23}=c_{24}$, $c_{25}=c_{26}$,
and, finally, 
\bea
c_{mn,ij}(\vec{k}) = c_{ij,mn}^*(\vec{k})
\eea
gives $c_4=c_5$. 

Assembling all this information, 
we find that the general 4th-order correlation tensor only depends on 7 scalar functions:
\bea
\nonumber
c_{ij,mn}(\vec{k}) &=& c_1(k)\delta_{ij}\delta_{mn} + 
c_2(k)\left(\delta_{im}\delta_{jn} + \delta_{in}\delta_{jm}\right)\\
&&+\,\, c_4(k)\left(\delta_{ij}\kk_m\kk_n + \delta_{mn}\kk_i\kk_j\right)
\nonumber\\ 
&&+\,\, c_6(k)\left(\delta_{im}\kk_j\kk_n + \delta_{in}\kk_j\kk_m 
+ \delta_{jm}\kk_i\kk_n + \delta_{jn}\kk_i\kk_m\right)
\nonumber\\
&&+\,\, c_{10}(k)\kk_i\kk_j\kk_m\kk_n
\nonumber\\
&&+\,\, i c_{17}(k)\left(\eps_{imp}\kk_p\delta_{jn} + \eps_{inp}\kk_p\delta_{jm}\right.
\nonumber\\
&&\qquad\qquad +\left. \eps_{jmp}\kk_p\delta_{in} + \eps_{jnp}\kk_p\delta_{im}\right)
\nonumber\\
&&+\,\, i c_{23}(k)\left(\eps_{imp}\kk_p\kk_j\kk_n + \eps_{inp}\kk_p\kk_j\kk_m\right.
\nonumber\\
&&\qquad\qquad +\left. \eps_{jmp}\kk_p\kk_i\kk_n + \eps_{jnp}\kk_p\kk_i\kk_m\right).
\label{eq::c4gen}
\eea

\subsection{Observables in the Case of Zero Mean Field}
\label{ap::Obs}

Let us first examine the case $\BB=0$, so we are only concerned with the 
4th-order statistics. We will need explicit expressions for the 
coordinate-dependent components of the tensor $c_{ij,mn}$ in terms 
of the coordinate-invariant functions $c_1$, $c_2$, $c_4$, $c_6$, $c_{10}$, 
$c_{17}$ and $c_{23}$ [\eqref{eq::c4gen}]. As the polarized emission data 
on the magnetic field arrives in the form of line-of-sight integrals (\secref{sec::SC}), 
we have to set $k_z=0$ everywhere---no information on the field variation 
in this direction is available. However, because of the assumed isotropy, 
the dependence of the invariant scalar functions on $k_\perp=|\vkp|$ 
contains the same information as their dependence on $k=|\vec{k}|$. 
Let us denote by $\phi$ the angle between $\vkp$ and the $x$ axis.
This means that we set $\vec{\kk} = (\cos\phi,\sin\phi,0)$. Then 
the components perpendicular to the line of sight are 
\bea
\nonumber
c_{xx,xx}(\vkp) &=& c_1 + 2c_2 + \left(2c_4 + 4c_6\right)\cos^2\phi + c_{10}\cos^4\phi,\\
\nonumber
c_{yy,yy}(\vkp) &=& c_1 + 2c_2 + \left(2c_4 + 4c_6\right)\sin^2\phi + c_{10}\sin^4\phi,\\ 
\nonumber
c_{xx,yy}(\vkp) &=& c_1 + c_4 + c_{10}\sin^2\phi\cos^2\phi,\\
\nonumber
c_{xy,xy}(\vkp) &=& c_2 + c_6 + c_{10}\sin^2\phi\cos^2\phi,\\
\nonumber
c_{xx,xy}(\vkp) &=& (c_4 + 2c_6)\cos\phi\sin\phi + c_{10}\cos^3\phi\sin\phi,\\
c_{yy,xy}(\vkp) &=& (c_4 + 2c_6)\cos\phi\sin\phi + c_{10}\cos\phi\sin^3\phi.
\label{eq::cxy}
\eea
These are the only components of $c_{ij,mn}$ that are directly sampled 
by the polarized emission. 
The components parallel to the line of sight are
\bea
\nonumber
c_{zz,zz}(\vkp) &=& c_1 + 2c_2,\\
\nonumber
c_{xx,zz}(\vkp) &=& c_1 + c_4\cos^2\phi,\\
\nonumber
c_{yy,zz}(\vkp) &=& c_1 + c_4\sin^2\phi,\\
\nonumber
c_{xz,xz}(\vkp) &=& c_2 + c_6\cos^2\phi,\\ 
\nonumber
c_{yz,yz}(\vkp) &=& c_2 + c_6\sin^2\phi,\\ 
c_{xz,yz}(\vkp) &=& c_6\sin\phi\cos\phi.
\label{eq::cz}
\eea
Information about these components can only be obtained by relying on 
the isotropy assumption as they are expressed in terms of the same invariant 
scalar functions as the perpendicular components. 
Note that setting $k_z=0$ has led to all information being lost about 
the mirror-asymmetric part of the tensor, so no quantity involving $c_{17}$ 
or $c_{23}$ can ever be reconstructed from polarized emission.  

\subsubsection{Stokes Correlators}
\label{ap::SCiso}

Using \eqref{eq::cxy} and the expressions for the Stokes correlators given 
by \eqref{eq::SC}, we get
\bea
\nonumber
\Sigma_{II}(\vkp) &=& 4\left(c_1 + c_2 + c_4 + c_6\right) + c_{10},\\
\nonumber
\Sigma_{QQ}(\vkp) &=& 4\left(c_2 + c_6\right) + c_{10}\cos^2 2\phi,\\
\nonumber
\Sigma_{UU}(\vkp) &=& 4\left(c_2 + c_6\right) + c_{10}\sin^2 2\phi,\\
\nonumber
\Sigma_{IQ}(\vkp) &=& (2c_4 + 4c_6 + c_{10})\cos2\phi,\\
\nonumber
\Sigma_{IU}(\vkp) &=& \left(2c_4 + 4c_6 + c_{10}\right)\sin2\phi,\\
\Sigma_{QU}(\vkp) &=& c_{10}\sin2\phi\cos2\phi,
\label{eq::SCiso}
\eea
Note that $\Sigma_{IQ}$ and $\Sigma_{IU}$ contain the same information 
and so do $\Sigma_{QU}$ and $\Sigma_{QQ}-\Sigma_{UU}$. 
The relations between them follow immediately from \eqref{eq::SCiso} 
and are given by \eqref{eq::IU_QU_rlns}. 

Thus, only 4 of the Stokes correlators are independent: $\Sigma_{II}$, 
two of $\Sigma_{QQ}$, $\Sigma_{UU}$, $\Sigma_{QU}$ and one of 
$\Sigma_{IQ}$, $\Sigma_{IU}$. 
We see that in \eqref{eq::SCiso}, these 4 independent observables are expressed 
in terms of 5 invariant scalar functions, $c_1$, $c_2$, $c_4$, $c_6$ and $c_{10}$, 
which cannot, therefore, all be reconstructed from polarized emission data 
even if isotropy is assumed (as we explained earlier, two other functions, 
$c_{17}$ and $c_{23}$, which complete the full set of 7 alluded to in 
\secref{sec::SC}, can never be known from polarized-emission data). 

The relations between the Stokes correlators and the invariant scalar functions 
given by \eqref{eq::SCiso} contain angular dependence. It will be convenient for 
practical calculations to express all observables in terms of angle averages 
(i.e., averages over all orientations of $\vkp$):
\bea
\nonumber
\Sigma_1(k)&=&{1\over2\pi}\int d\phi\,\Sigma_{II} = 4\left(c_1 + c_2 + c_4 + c_6\right) + c_{10},\\ 
\nonumber
\Sigma_2(k)&=&{1\over2\pi}\int d\phi\left(\Sigma_{QQ} + \Sigma_{UU}\right) 
= 8\left(c_2+c_6\right) + c_{10},\\
\nonumber
\Sigma_3(k)&=&{1\over2\pi}\int d\phi\left[\left(\Sigma_{QQ}-\Sigma_{UU}\right)\cos4\phi 
+ 2\Sigma_{QU}\sin4\phi\right] = c_{10},\\
\nonumber
\Sigma_4(k)&=&{1\over2\pi}\int d\phi\left(\Sigma_{IQ}\cos2\phi + \Sigma_{IU}\sin2\phi\right) 
= 2c_4 + 4c_6 + c_{10}.\\
\label{eq::SCavg}
\eea
These formulae again give us 4 independent observable scalar functions, but 
now the quality of the statistics should be improved by the angle 
averaging. We will find that it is most convenient for practical 
calculations to use $\Sigma_1$, \dots, $\Sigma_4$ as the basic set 
of observables (see \secref{ap::GenObs}). 

\subsubsection{General Observable Quantities}
\label{ap::GenObs}

Thus, only some 4th-order statistical quantities are observable. 
It is not hard to work out the condition for them to be so.
First, as we already explained in our discussion of \eqref{eq::cxy}, 
Stokes correlators carry no information about 
anything that involves the measures of mirror asymmetry of the field 
$c_{17}$ and $c_{23}$. Let us then restrict our attention to quantities 
that contain only the remaining 5 invariant scalar functions that determine 
the 4th-order two-point statistics [\eqref{eq::c4gen}].
In general, we would be looking for a scalar function that has the form
\bea
\Phi(k) = f_1 c_1(k) + f_2 c_2(k) + f_4 c_4(k) + f_6 c_6(k) + f_{10} c_{10}(k), 
\label{eq::Phi}
\eea
where $f_1$, \dots, $f_{10}$ are some known coefficients, which can be functions of $k$. 
Let us try to express this quantity in terms of Stokes correlators: 
this amounts to finding coefficients $\alpha$, $\beta$, $\gamma$, $\delta$, 
which can be functions of $k$ and $\phi$, such that 
\bea
\Phi(k) = {1\over4}\left(\alpha\Sigma_{II} + \beta\Sigma_{QQ} + \gamma\Sigma_{UU}
+ \delta\,{\Sigma_{IQ}\over\cos2\phi}\right). 
\label{eq::PhiS}
\eea
Using \eqref{eq::SCiso}, we get 
\bea
\nonumber
\Phi(k) &=& \alpha c_1 + \left(\alpha + \beta + \gamma\right) c_2 
+ \left(\alpha + {\delta\over2}\right) c_4 
+ \left(\alpha + \beta + \gamma + \delta\right) c_6\\ 
&&+\,\, {1\over4}\left(\alpha + \beta\cos^22\phi + \gamma\sin^22\phi + \delta\right)c_{10}.
\eea
Comparing this with \eqref{eq::Phi}, we get 
\bea
\nonumber
\alpha &=& f_1,\\
\nonumber
\beta &=& {1\over2}\left(f_2-f_1 + {3f_1 - f_2 - 4f_4 + 8f_{10}\over\cos4\phi}\right),\\
\nonumber
\gamma &=& {1\over2}\left(f_2-f_1 - {3f_1 - f_2 - 4f_4 + 8f_{10}\over\cos4\phi}\right),\\
\delta &=& 2\left(f_4-f_1\right) = f_6-f_2.
\label{eq::solve}
\eea
The last formula gives two expressions for $\delta$. 
In general, they do not have to be compatible and if they are 
not, the quantity $\Phi(k)$ cannot be expressed in terms of Stokes correlators. 
Thus, we have derived a simple criterion: 
only those quantities $\Phi(k)$ given by \eqref{eq::Phi} 
are observable for which 
\bea
2f_1 - f_2 - 2f_4 + f_6 = 0. 
\label{eq::crit}
\eea

If \eqref{eq::crit} is satisfied, \eqref{eq::solve} and \eqref{eq::PhiS} give 
us a practical method for calculating $\Phi$. As any interesting physical 
quantity $\Phi$ has to be independent of the angle $\phi$ between the wave vector 
$\vkp$ and the $x$ axis of the coordinate system in which the Stokes parameters 
are measured, we are allowed to average over~$\phi$:
\bea
\Phi(k) = {1\over8\pi}\int_0^{2\pi}d\phi W(\phi)
\left[\alpha\Sigma_{II} + \beta\Sigma_{QQ} + \gamma\Sigma_{UU}
+ \delta\,{\Sigma_{IQ}\over\cos2\phi}\right],
\label{eq::ang_avg}
\eea
where $W(\phi)$ is some weight function, which must satisfy  
$(1/2\pi)\int_0^{2\pi}d\phi W(\phi)=1$. 
The weighting does not theoretically affect the result, so $W(\phi)$ 
can be chosen arbitrarily. 

The possibility of angle averaging with a weight function and a certain redundancy 
of information available from the Stokes correlators, as expressed by \eqref{eq::IU_QU_rlns}, 
mean that there are, in general, many different ways of reconstructing 
observable quantities. In theory they are all equivalent, but in practice 
one has to choose one with the aim of reducing noise and offsetting the potentially 
detrimental effect of singularities in the coefficients associated with factors of 
$1/\cos2\phi$ and $1/\cos4\phi$. 

One method, which we have found to be quite effective, 
of avoiding this problem is to pick as our basic set of 4 observable scalar functions 
not the Stokes correlators themselves but the combinations of their angle averages 
given by \eqref{eq::SCavg}. Repeating the procedure we have just followed, 
we seek $\Phi(k)$ in the form
\bea
\Phi(k) = {1\over4}\left(\alpha_1\Sigma_1 + \alpha_2\Sigma_2 + \alpha_3\Sigma_3 + \alpha_4\Sigma_4\right),
\label{eq::PhiSavg}
\eea 
where the coefficients $\alpha_1$, \dots, $\alpha_4$ are now functions of $k$ only
[there is no longer any angular dependence on either side of \eqref{eq::PhiSavg}].  
Using \eqref{eq::SCavg}, this becomes
\bea
\nonumber
\Phi(k) \!\!\!\!\! &=& \!\!\!\!\! 
\alpha_1 c_1 + \left(\alpha_1 + 2\alpha_2\right) c_2 
+ \left(\alpha_1 + {\alpha_4\over2}\right)c_4 
+ \left(\alpha_1 + 2\alpha_2 + \alpha_4\right)c_6\\
&&+\,\, {1\over4}\left(\alpha_1 + \alpha_2 + \alpha_3 + \alpha_4\right)c_{10}.
\eea
Comparing this expression with \eqref{eq::Phi} as before, we get 
\bea
\nonumber
\alpha_1 &=& f_1,\\
\nonumber
\alpha_2 &=& {1\over2}\left(f_2-f_1\right),\\
\nonumber
\alpha_3 &=& {1\over2}\left(3f_1 - f_2 - 4f_4 + 8 f_{10}\right),\\
\alpha_4 &=& 2(f_4-f_1) = f_6-f_2,
\label{eq::solve_avg}
\eea
and the observability criterion is again given by \eqref{eq::crit}.
\eqref{eq::PhiSavg} together with \eqref{eq::SCavg} and \eqref{eq::solve_avg} 
give another expression for a general observable $\Phi(k)$, defined 
by \eqref{eq::Phi} and subject to the constraint \eqref{eq::crit}.  

\subsubsection{Tension-Force Power Spectrum}
\label{ap::TenForce}

In \secref{sec::TenForce}, we split the tension-force power spectrum 
into the directly observable part [\eqref{eq::Phi1}], which could be 
recovered from the Stokes correlators without any assumptions, 
and the non-directly-observable part $\Phi_2$ [\eqref{eq::Phi2}], 
which could only be reconstructed using some assumed symmetries of the 
4th-order correlation tensor. Assuming isotropy, 
we infer from \eqref{eq::cz} 
\bea
\Phi_2 = k^2(c_2 + c_6). 
\eea
Using \eqref{eq::cxy} to express $c_2+c_6$ in terms of the perpendicular 
components of the tensor $c_{ij,mn}$, we immediately recover \eqref{eq::Phi2iso}
and the rest follows as explained in \secref{sec::TenForce}. 

This was an {\em ad hoc} derivation specific to the tension-force 
power spectrum. Let us now demonstrate how the general method laid out 
in \secref{ap::GenObs} works for this quantity. 

Substituting \eqref{eq::c4gen} into \eqref{eq::MTFT}, we get 
\bea
\Phi(k) = k^2 \left[c_1(k) + 4c_2(k) + 2c_4(k) + 6c_6(k) + c_{10}(k)\right],
\label{eq::Phi_from_c}
\eea
a particular case of \eqref{eq::Phi}.  
The observability criterion given by \eqref{eq::crit} is satisfied, so, 
using \eqref{eq::PhiS} and \eqref{eq::solve}, we obtain
\bea
\Phi(k) &=& {1\over4} k^2 \left[\Sigma_{II} 
+ {1\over2}\left(3 - {1\over\cos4\phi}\right)\Sigma_{QQ}
\right.\nonumber\\
&&+\left. {1\over2}\left(3 + {1\over\cos4\phi}\right)\Sigma_{UU}
+ {2\over\cos 2\phi}\Sigma_{IQ}\right],
\label{eq::Phi_iso}
\eea 
which can be angle-averaged with some weight function according to \eqref{eq::ang_avg}. 

An alternative expression in terms of averaged Stokes correlators 
follows from \eqref{eq::PhiSavg} and \eqref{eq::solve_avg}:
\bea
\Phi(k) = {1\over4}k^2\left(\Sigma_1 + {3\over2}\Sigma_2 - {1\over2}\Sigma_3 + 2\Sigma_4\right).
\label{eq::Phi_from_Sigmas}
\eea
Substituting for $\Sigma_1$, \dots, $\Sigma_4$ 
from \eqref{eq::SCavg}, we arrive at 
\bea
\nonumber
\Phi(k) \!\!\!\!\! &=& \!\!\!\!\!
{k^2\over8\pi}\int_0^{2\pi} d\phi \biggl[\Sigma_{II} 
+ 2\left(\Sigma_{IQ}\cos2\phi + \Sigma_{IU}\sin2\phi\right)
- \Sigma_{QU}\sin4\phi
\biggr.\\
&&+\,\,\biggl. {1\over2}\left(3-\cos4\phi\right)\Sigma_{QQ}
+ {1\over2}\left(3+\cos4\phi\right)\Sigma_{UU}\biggr].
\label{eq::Phi_iso_avg}
\eea
Our final formula for the tension-force power spectrum, \eqref{eq::Tk_obs}, 
follows from \eqref{eq::Phi_iso_avg} upon multiplication by $4\pi k^2$
(the wave-vector-space volume factor). 
Note that the integrand in \eqref{eq::Phi_iso_avg} 
reduces back to \eqref{eq::Phi_iso} if we make use 
of \eqref{eq::IU_QU_rlns}, but the advantage of \eqref{eq::Phi_iso_avg} 
is that it does not contain any singular coefficients.

\subsection{Observables in the Case of Weak Mean Field}
\label{ap::Obs_wmf}

If a weak mean field is present, we proceed analogously to 
\secref{ap::Obs}. It is understood that the mean field is 
sufficiently weak so as not to break the isotropy of the fluctuating 
part of the field. Then, using \eqref{eq::Ckgeneral}, \eqref{eq::c2gen} 
and \eqref{eq::c3gen} and 
setting $k_z=0$ to express the line-of-sight integrals, we find
\bea
\nonumber
C_{xx,xx}(\vkp) &=& c_{xx,xx}(\vkp) + 2\BB_x^2\sin^2\phi\,m_1,\\ 
\nonumber 
C_{yy,yy}(\vkp) &=& c_{yy,yy}(\vkp) + 2\BB_y^2\cos^2\phi\,m_1,\\
\nonumber
C_{xx,yy}(\vkp) &=& c_{xx,yy}(\vkp) - 2\BB_x\BB_y\sin\phi\cos\phi\,m_1\\ 
\nonumber
&&+\,\, 4i\left(\BB_x\sin\phi + \BB_y\cos\phi\right)\sin\phi\cos\phi\,a_2,\\
\nonumber
C_{xy,xy}(\vkp) &=& c_{xy,xy}(\vkp) + {1\over2}\left(\BB_x\cos\phi - \BB_y\sin\phi\right)^2 m_1,\\
\nonumber
C_{xx,xy}(\vkp) &=& c_{xx,xy}(\vkp) - \BB_x\sin\phi\left(\BB_x\cos\phi - \BB_y\sin\phi\right) m_1\\
\nonumber
&&-\,\, 2i\left(\BB_x\sin\phi - \BB_y\cos\phi\right)\sin^2\phi\,a_2,\\
\nonumber
C_{yy,xy}(\vkp) &=& c_{yy,xy}(\vkp) + \BB_y\cos\phi\left(\BB_x\cos\phi - \BB_y\sin\phi\right) m_1\\
&&+\,\, 2i\left(\BB_x\sin\phi - \BB_y\cos\phi\right)\cos^2\phi\,a_2,
\label{eq::Cxy_wmf}
\eea
where the components of $c_{ij,mn}(\vkp)$ are given by \eqref{eq::cxy} and \eqref{eq::cz}. 

\subsubsection{Stokes Correlators}

Using \eqref{eq::Cxy_wmf} and \eqref{eq::SC}, we find that the Stokes 
correlators~are
\bea
\nonumber
\Sigma_{II}(\vkp) &=& 2\left(\BB_x\sin\phi - \BB_y\cos\phi\right)^2 m_1 + {\rm 4th~order},\\
\nonumber
\Sigma_{QQ}(\vkp) &=& 2\left(\BB_x\sin\phi + \BB_y\cos\phi\right)^2 m_1 + {\rm 4th~order},\\
\nonumber
\Sigma_{UU}(\vkp) &=& 2\left(\BB_x\cos\phi - \BB_y\sin\phi\right)^2 m_1 + {\rm 4th~order},\\
\nonumber
\Sigma_{IQ}(\vkp) &=& 2\left(\BB_x^2\sin^2\phi - \BB_y^2\cos^2\phi\right) m_1\\
\nonumber 
&&-\,\, 4i\left(\BB_x\sin\phi - \BB_y\cos\phi\right)\sin2\phi\,a_2 + {\rm 4th~order},\\
\nonumber
\Sigma_{IU}(\vkp) &=& -\left[\left(\BB_x^2+\BB_y^2\right)\sin2\phi - 2\BB_x\BB_y\right] m_1\\ 
\nonumber
&&+\,\, 4i\left(\BB_x\sin\phi - \BB_y\cos\phi\right)\cos2\phi\,a_2 + {\rm 4th~order},\\
\nonumber
\Sigma_{QU}(\vkp) &=& -\left[\left(\BB_x^2-\BB_y^2\right)\sin2\phi + 2\BB_x\BB_y\cos2\phi\right] m_1\\
&&-\,\, 4i\left(\BB_x\sin\phi - \BB_y\cos\phi\right) a_2 + {\rm 4th~order},
\label{eq::SCiso_wmf}
\eea 
where the 4th-order parts of the correlators are given by \eqref{eq::SCiso}. 

Thus, the Stokes correlators now contain not just the 4th-order statistics 
but also some information about the second- and 3rd-order correlation 
functions of the field, namely, the magnetic-field power spectrum $m_1(k)$ [see \eqref{eq::c2gen}] 
and the 3rd-order correlation function $a_2(k)$ [\eqref{eq::c3gen}]. 
We are not particularly interested in $a_2$ and notice that all terms containing it 
can be eliminated from \eqref{eq::SCiso_wmf} simply by taking the real part of the 
Stokes correlators $\Sigma_{IQ}$, $\Sigma_{IU}$ and $\Sigma_{QU}$. 
We will now isolate the second- and 4th-order contributions to the Stokes correlators
and calculate the power spectra of the magnetic field and of the tension force. 

\subsubsection{Magnetic-Field Power Spectrum}
\label{ap::MFSp}

There are several formulae that allow one to distill $m_1$ from the Stokes correlators. They are 
all derived by assembling appropriate linear combinations of the correlators and 
angle averaging. The simplest such formulae are found by noticing that the angle 
averages of the 4th-order parts of $\Sigma_{IQ}$, $\Sigma_{IU}$ 
and of $\Sigma_{QQ}-\Sigma_{UU}$ vanish [see \eqref{eq::SCiso}],~so 
\bea
\nonumber
&&\!\!\!\!\!\! {1\over2\pi}\int_0^{2\pi} d\phi\,{\rm Re}\Sigma_{IQ} = \left(\BB_x^2 - \BB_y^2\right) m_1,\\ 
\nonumber
&&\!\!\!\!\!\! {1\over2\pi}\int_0^{2\pi} d\phi\,{\rm Re}\Sigma_{IU} = 2\BB_x\BB_y m_1,\\
&&\!\!\!\!\!\! {1\over2\pi}\int_0^{2\pi} d\phi \left(\Sigma_{QQ} - \Sigma_{UU}\right) = 8\BB_x\BB_y m_1,
\label{eq::m1_avg}
\eea
The standard one-dimensional 
magnetic-field power spectrum is defined with an additional wave-number-space 
volume factor of $4\pi k^2$ [\eqref{eq::Mdef}]---the resulting expressions for it 
are given in \eqref{eq::Mk_obs}. 

It is also possible to construct formulae that formally do not require angle averaging at all: 
using the fact that for certain combinations of the Stokes correlators the 4th-order 
contributions must vanish [assuming isotropy; see \eqref{eq::IU_QU_rlns}], we find 
\bea
\nonumber
m_1 &=& {{\rm Re}\left(\Sigma_{IQ}\sin2\phi - \Sigma_{IU}\cos2\phi\right)\over 
\left(\BB_x^2-\BB_y^2\right)\sin2\phi - 2\BB_x\BB_y\cos2\phi},\\
m_1 &=& -{\left(\Sigma_{QQ}-\Sigma_{UU}\right)\sin4\phi - 2{\rm Re}\Sigma_{QU}\cos4\phi\over
2\left[\left(\BB_x^2-\BB_y^2\right)\sin2\phi - 2\BB_x\BB_y\cos2\phi\right]}.
\label{eq::m1_direct}
\eea
Since $m_1$ is independent of the orientation of the wave vector, 
these expressions can be averaged over the angle $\phi$ with arbitrary weight functions. 
Note that in order to compute the shape of the spectrum from \eqref{eq::m1_direct}, 
we do not need to know the magnitude of the mean field but we do need its 
orientation (in the plane perpendicular to the line of sight). 

In practice, we expect the formulae given by 
\eqref{eq::m1_avg} to work better than those given by \eqref{eq::m1_direct} 
because the latter rely on exact cancellations that are probably not going 
to happen with very high precision in realistic situations. 
Even moderate errors in cancelling the 4th-order correlators could then easily overwhelm the 
second-order ones: indeed, the terms in \eqref{eq::SCiso_wmf} involving the mean 
field are small because the mean field was assumed to be weak: $\BB^2\ll\la b^2\ra$, 
so $\BB^2 m_1\ll c_{ij,mn}$. In \eqref{eq::m1_avg}, the cancellation of the 
4th-order correlators comes from angle averaging and there is hope that 
$m_1(k)$ could be recovered (but see the caveat at the end of \secref{sec::weakmf}).  

\subsubsection{Magnitude and Orientation of the Mean Field}
\label{ap::mf}

The orientation of the mean field (or, rather, its projection on the plane perpendicular 
to the line of sight) can be easily determined from the Stokes parameters 
themselves: the angle $\Theta$ between the mean field and the $x$ axis satisfies  
\bea
\tan 2\Theta = {2\BB_x\BB_y\over\BB_x^2 - \BB_y^2} = {\la U\ra\over\la Q\ra}. 
\label{eq::Theta}
\eea
Note that $\tan2\Theta$ tells us the orientation but not the sign of the mean field. 
This angle can also be determined from the Stokes correlators via \eqref{eq::m1_avg}:
\bea
\tan 2\Theta = {\int_0^{2\pi}d\phi\,{\rm Re}\Sigma_{IU}\over\int_0^{2\pi}d\phi\,{\rm Re}\Sigma_{IQ}}
= {\int_0^{2\pi}d\phi\left(\Sigma_{QQ}-\Sigma_{UU}\right)\over4\int_0^{2\pi}d\phi\,{\rm Re}\Sigma_{IQ}},
\label{eq::ThetaSC}
\eea
but the validity of these formulae, unlike that of \eqref{eq::Theta}, is predicated on 
assuming the statistical isotropy of the fluctuating field. 
If this assumption is well satisfied, the value of $\tan 2\Theta$ obtained 
from \eqref{eq::ThetaSC} should be independent of $k$. 

The magnitude of the mean field is a slightly more complicated quantity to determine. 
From the total emission intensity [see \eqref{eq::StoPar}], 
\bea
\la I\ra = \BB_\perp^2 + \la b_x^2\ra + \la b_y^2\ra = \BB_\perp^2 + {2\over3}\la b^2\ra,
\label{eq::Iavg}
\eea
where $\BB_\perp^2=\BB_x^2+\BB_y^2$ 
and the last expression follows from assuming the statistical isotropy 
of the fluctuating field $\vec{b}$. Thus, from averaging $I$, we can calculate the total 
energy density of the magnetic field but not individually the magnitudes of the mean 
and fluctuating fields. On the other hand, once we know $\Theta$, 
we can find $\BB_\perp^2 m_1$ from \eqref{eq::m1_avg} or \eqref{eq::m1_direct}. 
Let us integrate this 
quantity over all wavenumbers and denote the result by $A$: 
\bea
\BB_\perp^2\int_0^\infty dk\, 4\pi k^2 m_1(k) = \BB_\perp^2 \la b^2\ra = A.
\eea
Then $\la b^2\ra = A/\BB_\perp^2$ and substituting this into \eqref{eq::Iavg}, we arrive 
at a biquadratic equation for $\BB_\perp$, whose solution is
\bea
\BB_\perp^2 = {1\over2}\left(\la I\ra - \sqrt{\la I\ra^2 - {8\over3}\,A}\right) 
\simeq {2\over3} {A\over\la I\ra}.
\label{eq::BBperp}
\eea
We have chosen the ``$-$'' root because we are assuming $\BB^2\ll\la b^2\ra$ 
(weak mean field). While we do not really need to know $\BB_\perp^2$ to disentangle 
the second- and 4th-order statistics in \eqref{eq::SCiso_wmf}, we can use 
\eqref{eq::BBperp} to check that the mean field really is weak:
\bea
\BB_\perp^2\ll \la I\ra. 
\eea

\subsubsection{Fourth-Order Quantities}

Now that we know the mean field and $m_1(k)$, we can use this information  
in \eqref{eq::SCiso_wmf} to isolate 
the 4th-order statistics in the Stokes correlators and then proceed 
to calculating all observable 4th-order quantities in the same way it was done 
in \secref{ap::GenObs}. 
In general, this involves subtracting from the (real part of) the 
Stokes correlators the terms that contain $m_1$ [\eqref{eq::SCiso_wmf}] 
so that only the 4th-order contributions [\eqref{eq::SCiso}] remain. 
Doing this requires knowing $\BB_\perp^2m_1(k)$ (see \secref{ap::MFSp}) and 
the orientation of the mean field [\eqref{eq::Theta}].
Subtracting the second-order contributions from the 
the averaged Stokes correlators introduced in \eqref{eq::SCavg} 
is a particularly simple operation: 
substituting from \eqref{eq::SCiso_wmf} into \eqref{eq::SCavg} and carrying 
out the angle averages, we get
\bea
\nonumber
\Sigma_1(k) &=& \BB_\perp^2 m_1 + {\rm 4th~order},\\
\nonumber
\Sigma_2(k) &=& 2\BB_\perp^2 m_1 + {\rm 4th~order},\\
\nonumber
\Sigma_3(k) &=& {\rm 4th~order},\\
\Sigma_4(k) &=& -{3\over 4}\BB_\perp^2 m_1 + {\rm 4th~order},
\label{eq::SCavg_wmf}
\eea
where the 4th-order parts are given by \eqref{eq::SCavg} and real part
of the Stokes correlators is taken everywhere to eliminate the 
contributions from the 3rd-order statistics. 

\subsubsection{Tension-Force Power Spectrum}
\label{ap::TenForce_wmf}
 
To find the tension-force power spectrum when $\BB\neq0$, 
substitute \eqref{eq::Ckgeneral} into \eqref{eq::MTFT} and use 
the solenoidality of the magnetic field [\eqref{eq::sol2} and \eqref{eq::sol3}]:  
\bea
\Phi = \BB_j\BB_n k_j k_n c_{i,i} 
+ \BB_n k_n k_j \left(c_{ij,i} + c_{ij,i}^*\right)
+ k_jk_n c_{ij,in}.
\eea
From \eqref{eq::c3gen}, we see that the second term vanishes. 
Using \eqref{eq::c2gen} to express the first term in terms of the 
magnetic-field power spectrum, 
setting $\vec{k} = k_\perp(\cos\phi,\sin\phi,0)$ (line-of-sight integral)
and angle-averaging over $\phi$, we get 
\bea
\Phi(k) = {1\over2}\,k^2\BB_\perp^2 m_1(k) + {\rm 4th~order}, 
\eea
where the 4th-order part is given by \eqref{eq::Phi_from_c} 
and is calculated from Stokes correlators in the same way 
as in \secref{ap::TenForce}. 
Namely, using \eqref{eq::Phi_from_Sigmas} and \eqref{eq::SCavg_wmf}, we get 
\bea
\Phi(k) = {1\over4}k^2\left(\Sigma_1 + {3\over2}\Sigma_2 - {1\over2}\Sigma_3 + 2\Sigma_4
- {1\over2}\,\BB_\perp^2 m_1\right),
\eea
where $\Sigma_1$, \dots, $\Sigma_4$ are defined in \eqref{eq::SCavg} (with real 
parts taken of the Stokes correlators) and $\BB_\perp^2m_1$ is calculated 
via one of the formulae given in \secref{ap::MFSp}. 
\eqref{eq::Tk_obs_wmf} follows upon multiplication by the 
wave-number-space volume factor $4\pi k^2$. 

\bsp

\label{lastpage}

\end{document}